# 3D molecular representation learning for organic mixtures: viscosity and density prediction

Haicheng Qu[1+], Yanyi Su[2,3,+], Ning Wang[1], Shangqian Chen[3], Zhifeng Gao[3*], Jun Cheng[2,4,5*], Qi Ou[1*]

[1]Research Institute of Petroleum Processing, Sinopec Corporation, 18th Xueyuan Road, Haidian District, Beijing, China

[2]State Key Laboratory of Physical Chemistry of Solid Surface, College of Chemistry and Chemical Engineering, Xiamen University, Xiamen, China

[3]DP Technology, Beijing, China

[4]Laboratory of AI for Electrochemistry (AI4EC), Tan Kah Kee Innovation Laboratory (IKKEM), Xiamen, China

[5]Institute of Artificial Intelligence, Xiamen University, Xiamen, China

*Email: gaozf@dp.tech, chengjun@xmu.edu.cn, ouqi.ripp@sinopec.com

+These authors contributed equally to this work.

## Abstract

The viscosity and density of organic mixtures are essential properties for designing lubricants, solvents, and heat transfer fluids. In engineering practice, formulating a functional fluid requires understanding how these properties change with composition and temperature. However, exhaustive experimental characterization across the full parameter space is impractical due to the vast number of possible species and combinations. Here we introduce a mixture-aware 3D molecular representation learning strategy, built upon a pre-trained molecular encoder, that jointly encodes component structures, mole fractions, and temperature to achieve accurate predictions for organic mixtures. Fine-tuning on publicly available datasets covering a wide range of binary organic mixtures yields test-set $R^2$ values of 0.973 for dynamic viscosity and 0.996 for density, significantly outperforming traditional machine learning baselines. Beyond this overall accuracy, the model captures non-monotonic viscosity changes upon mixing, surpassing simple linear or logarithmic mixing rules. The architecture is extendable to ternary and multicomponent mixtures, as verified via preliminary experiments. Using this model, we quantitatively analyze how molecular structure—branching, cycloalkane, and aromatic rings—affects viscosity-temperature behavior, which benefits the design of lubricants with superior viscosity-temperature performance. Altogether, this work provides a practical, data-driven tool for mixture property prediction, accelerating the rational formulation of functional fluids in chemical engineering.

## Introduction

Dynamic viscosity and density are the key performance-defining properties of lubricants, solvents, and heat transfer fluids. Designing such high-performance fluids relies on organic mixtures whose dynamic viscosity and density must be precisely tuned across composition and temperature[1–5]. Viscosity exerts a dual influence: excessively high viscosity increases frictional energy loss, while insufficiently low viscosity fails to maintain a protective lubricant film, accelerating wear[6,7]. Density, in turn, governs volumetric metering and equipment sizing[8,9]. In engineering practice, however, exhaustively characterizing these properties over the full composition-temperature space is infeasible, not only because of the combinatorial explosion of possible pairs but also due to the strongly nonlinear mixing behaviors (arising from hydrogen bonding, dipole-dipole interactions, and dispersion forces[10–12]) that make simple interpolation unreliable.

Conventional approaches to obtaining mixture properties fall into three categories, each with inherent limitations. Experimental measurements, while accurate for a single sample, cannot keep pace with the vast design space of organic mixtures[13]. Molecular dynamics (MD) simulations offer a physics-based alternative, yet reliable viscosity extraction via the Green–Kubo relation requires extremely long trajectories and often suffers from slow convergence of the stress-tensor autocorrelation function[14,15]. Empirical quantitative structure-property relationship (QSPR) models[16] and traditional machine learning (ML) methods such as random forests, XGBoost, and support vector machines[17–20] provide faster predictions, but their performance is fundamentally constrained by hand-crafted molecular descriptors that cannot fully capture high-order intermolecular interactions in mixtures, especially when hydrogen bonding or non-ideal packing occurs[21–25].

Deep learning has enabled the development of powerful molecular representation learning techniques[26–32]. Graph neural networks (GNNs) operating on 2D molecular graphs and sequence-based models using SMILES strings have shown success in both pure-component property prediction[33,34] and binary mixtures[2], but effectively capturing composition-dependent cooperative interactions remains challenging for these models, suggesting room for improvement with richer molecular representations. More recently, the Uni-Mol framework[35,36] introduced a 3D Transformer pre-trained on 209 million molecular conformations, learning rich geometric and electronic features through self-supervised tasks. Uni-Mol has demonstrated superior performance for pure-component properties[37,38] and has since been extended to mixture property prediction in electrolyte formulation design[39]. That being said, its potential for predicting viscosity and density of general organic mixtures, including those exhibiting complex composition- and temperature-dependent behavior, has not been systematically explored.

Here we introduce Uni-Mix, a mixture-aware 3D molecular representation learning strategy built upon the Uni-Mol pre-trained encoder, that jointly encodes component structures, mole fractions, and temperature to achieve accurate predictions for organic mixtures. This work makes several key contributions to the field. Methodologically, we establish a practical pipeline that adapts a pre-trained 3D molecular representation model to mixture property prediction, achieving test-set $R^2$ values of 0.973 for viscosity and 0.996 for density of binary organic mixtures, which substantially outperforms conventional ML baselines[2,17,18]. Notably, the model captures non-monotonic viscosity-composition relationships that cannot be described by simple linear or logarithmic mixing rules. Moreover, the same architecture readily generalizes to ternary and multicomponent mixtures, making it applicable to training and prediction for such systems. Leveraging this model, we further analyze how molecular structural features—branching, cycloalkane rings, aromatic rings, and hydrogen-bonding capability—influence viscosity-temperature behavior and nonlinear composition dependence, providing practical guidelines for designing fluids with enhanced viscosity-temperature performance. Together, these results establish a data-driven, transferable, and physically interpretable framework for mixture property prediction, accelerating the rational formulation of functional fluids in chemical engineering.

# Methods

We adopt the pre-trained Uni-Mol framework as the base model architecture, which was pre-trained on 209 million small organic molecules via self-supervised 3D coordinate recovery and masked atom prediction tasks[35]. For each component in a mixture, a three-dimensional conformer is generated from its SMILES representation using RDKit[40] and fed into the Uni-Mol encoder to obtain a 512-dimensional molecular embedding. An additional 15-dimensional RDKit descriptor vector capturing molecular weight, LogP, polar surface area, hydrogen bonding capacity, ring counts, and key functional group counts is concatenated to form a 527-dimensional single-molecule representation. All components share the same encoder weights. The single-molecule representations are then processed by a mixture interaction module comprising multi-head self-attention layers followed by an aggregation operation, as illustrated in Figure 1. Mole fractions and temperature are injected via element-wise multiplication and sinusoidal positional encoding, respectively, yielding a composition- and temperature-aware mixture representation. Specifically, the temperature $T$ (in Kelvin) is encoded into a $d$-dimensional vector using sinusoidal positional encoding:

$$PE(T)_{2k} = \sin\left(\frac{T}{\omega^{2k/d}}\right), \quad PE(T)_{2k+1} = \cos\left(\frac{T}{\omega^{2k/d}}\right)$$

where $T$ is the temperature in Kelvin, $k$ is the dimension index ($k$ = 0, 1, ..., $d$/2 - 1), $d$ = 256 is the encoding dimension, and $\omega$ = 10000 is the base frequency. This parameter-free encoding projects the temperature into a high-dimensional space with multi-scale periodicity, providing richer input features than a single scalar without introducing additional trainable parameters. The final mixture representation is mapped to a two-dimensional output ($\mu_i$, $\log\sigma_i^2$) and trained with Gaussian negative log-likelihood (NLL) loss:

$$\mathcal{L} = \frac{1}{2N}\sum_{i=1}^{N}\left[e^{-\log\sigma_i^2}(y_i - \mu_i)^2 + \log\sigma_i^2\right]$$

where $N$ is the number of samples, $y_i$ the experimentally measured value for sample $i$, $\mu_i$ the predicted mean value for sample $i$, and $\log\sigma_i^2$ is the predicted log-variance for sample $i$, clamped to [-10, 10] for numerical stability. Compared with standard mean squared error loss, Gaussian NLL loss adaptively assigns lower penalty to samples with inherently higher prediction difficulty, preventing hard examples such as hydrogen-bonding systems from dominating gradient updates. To handle pure components and mixtures within a unified architecture, a masking mechanism automatically excludes absent components, allowing the model to degrade gracefully to single-molecule processing without separate branches. Pure-component data are included in training across all cross-validation folds but excluded from test sets, ensuring that generalization is evaluated exclusively on mixture data.

Property-specific fine-tuning is performed independently for viscosity and density using experimental data of binary organic mixtures from the Landolt–Börnstein database[41,42], randomly partitioned into training and test sets at an 8:2 ratio using a grouped split keyed on the molecular pair (SMILES_I, SMILES_II), so that all records sharing the same pair are assigned as a single unit, ensuring that no mixture pair in the test set appears during training. The size and range of each property dataset are summarized in Table 1. For dynamic viscosity, the target is $\log_{10}$-transformed prior to training, while density is used in its native linear scale. Accordingly, for viscosity, $R^2$ is evaluated in the $\log_{10}$-transformed space, whereas MAE and RMSE are reported in the native linear space (in the unit of cP) after back-transformation. For density, all metrics are in native linear space (g/cm$^3$). The three metrics— $R^2$, MAE, and RMSE—are defined as follows:

$$R^2 = 1 - \frac{\sum_{i=1}^{n}(y_i - \hat{y}_i)^2}{\sum_{i=1}^{n}(y_i - \bar{y})^2}$$

$$MAE = \frac{1}{n}\sum_{i=1}^{n} | y_i - \hat{y}_i |$$

$$RMSE = \sqrt{\frac{1}{n}\sum_{i=1}^{n}(y_i - \hat{y}_i)^2}$$

where n is the number of samples, $y_i$ is the experimentally measured value for sample $i$, $\hat{y}_i$ is the Uni-Mix predicted value for sample $i$, and $\bar{y}$ is the mean of the experimental data. Note that although we fine-tune and evaluate the model using binary mixture data, the architecture itself does not assume a fixed number of components. It can be directly applied to ternary and multicomponent systems by simply including additional component embeddings and their corresponding mole fractions as shown in Figure 1. For experimental validation, dynamic viscosity was measured using an Anton Paar SVM 3001 viscometer in accordance with ASTM D7042, and density was determined in accordance with ASTM D4052.

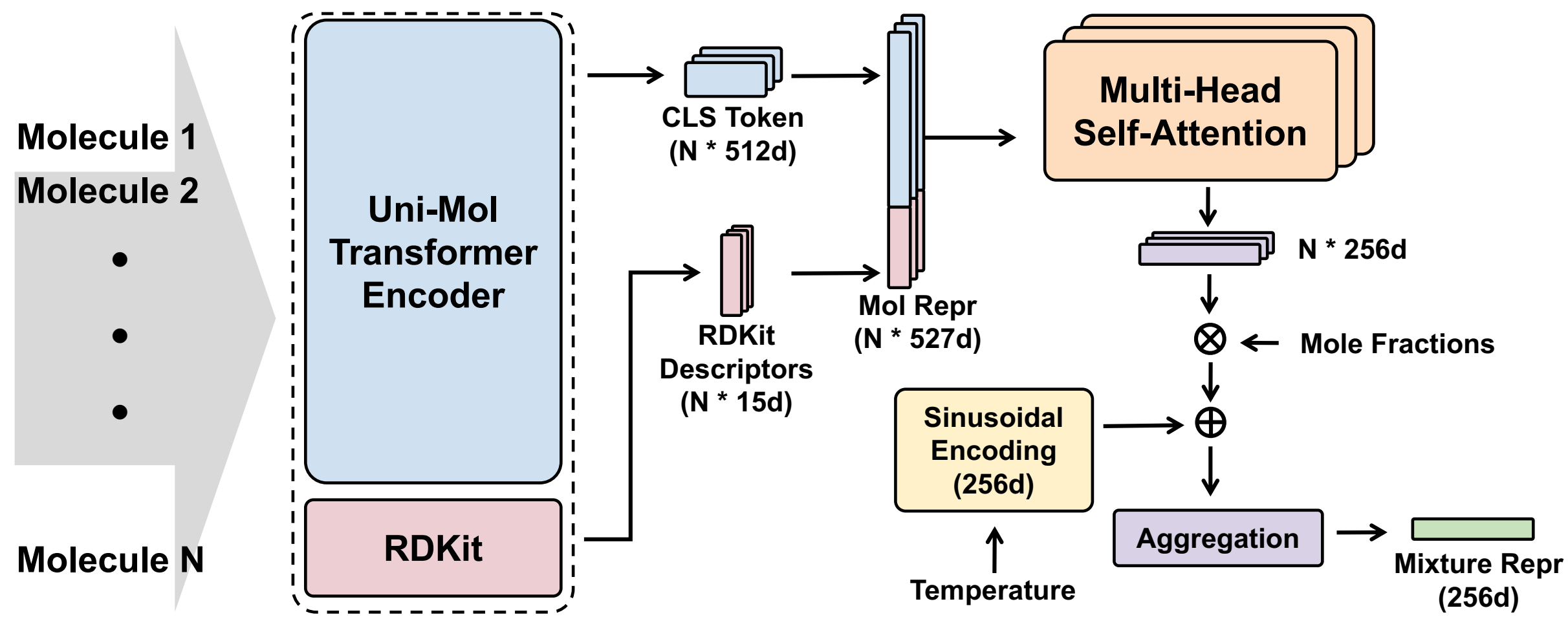

**Figure 1 | Schematic diagram of the Uni-Mix model pre-training and fine-tuning process for organic mixtures.**

**Table 1 | Dataset size and value ranges of the fine-tuning datasets used in this study, including dynamic viscosity η and density ρ of binary mixtures measured over the temperature range of 0–125 °C.**

| Property (0–125 °C) | Pure-component | Mixture | Total samples | Min. value | Max. value |
|---|---|---|---|---|---|
| Dynamic viscosity η | 5144 | 29,161 | 34,305 | 0.0503 | 60.81 |
| Density ρ | 3333 | 25,483 | 28,816 | 0.621 | 1.921 |

*Note: Units for η and ρ are cP and g/cm³, respectively. The dataset encompasses a broad range of organic compound classes including hydrocarbons, esters, anhydrides, alcohols, carboxylic acids, and other functionalized organics.*

# Results and Discussion

## Prediction of dynamic viscosity and density

We examine the capability of the fine-tuned model in predicting dynamic viscosity and density, two thermophysical properties of practical importance in applications ranging from lubricant formulation to chemical process design and solvent selection. Note that compared with density, viscosity is more sensitive to temperature and intermolecular interactions, and thus represents a more comprehensive prediction target[5,43]. We therefore carry out two independent fine-tuning tasks based on the pre-trained model for these two properties.

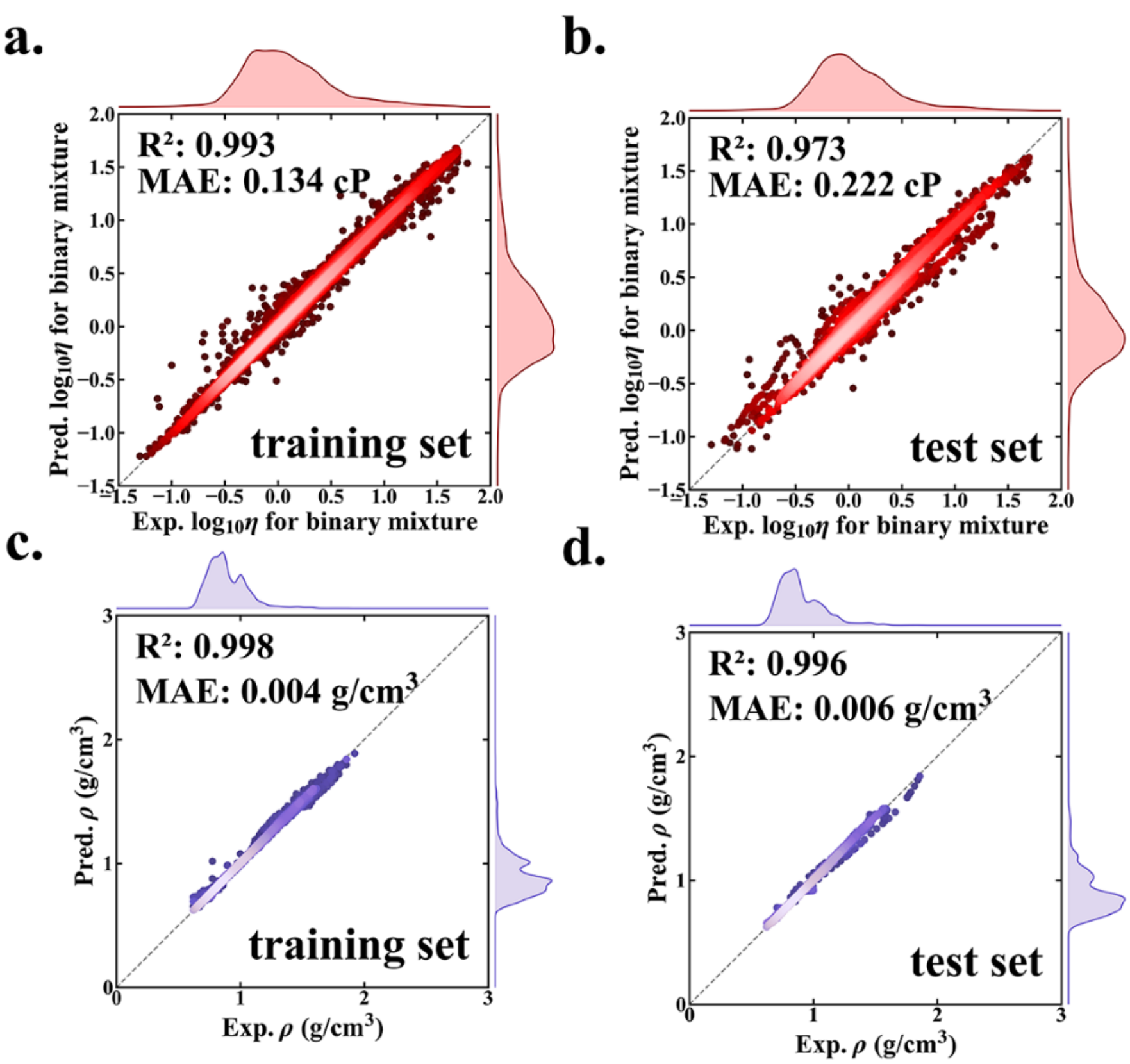

**Figure 2 | Performance of the Uni-Mix model in predicting dynamic viscosity and density of binary mixtures.** ***a*** Correlation between Uni-Mix predicted and experimental dynamic viscosity values for the training set ($R^2$ = 0.993, MAE = 0.134 cP). ***b*** Corresponding parity plot for the test set ($R^2$ = 0.973, MAE = 0.222 cP). ***c*** Correlation between Uni-Mix predicted and experimental density values for the training set ($R^2$ = 0.998, MAE = 0.004 g/cm$^3$). ***d*** Corresponding parity plot for the test set ($R^2$ = 0.996, MAE = 0.006 g/cm$^3$). Dashed lines indicate the ideal y = x correspondence.

As shown in Figure 2, our fine-tuned Uni-Mix model achieves high predictive accuracy for both properties, with density predictions (test $R^2$ = 0.996) outperforming those for dynamic viscosity (test $R^2$ = 0.973). This difference is attributable to the more tractable structure-property relationship of density, which is primarily governed by molecular mass and volume. The residual outliers in the viscosity predictions are concentrated in carboxylic acid-containing systems, where strong hydrogen-bonding interactions[23,44,45] lead to highly nonlinear viscosity-composition relationships that the current model struggles to capture. For density, the few outliers are confined to high-density regions where training data coverage is limited, suggesting that targeted data augmentation could further improve performance.

To further benchmark Uni-Mix fine-tuned results against conventional approaches, we conduct comparative experiments using three representative machine learning models, i.e., support vector machine (SVM), random forest (RF), and XGBoost. Each molecule was represented by a set of 2D physicochemical descriptors computed using RDKit, serving as input features for the three baseline models. Hyperparameters are optimized for each baseline model to ensure a fair comparison, as detailed in Tables S1–S4. The metrics for the test set are summarized in Table 2.

**Table 2 | Performance comparison of Uni-Mix (our work), SVM, XGBoost, and Random Forest for predicting dynamic viscosity η and density ρ of binary mixtures.**

| Metric | Model | η | ρ |
|---|---|---|---|
| ***R*²** | **Our work** | 0.973 | 0.996 |
| | **SVM** | 0.740 | 0.971 |
| | **XGBoost** | 0.955 | 0.984 |
| | **Random Forest** | 0.901 | 0.965 |
| **MAE** | **Our work** | 0.222 | 0.006 |
| | **SVM** | 0.671 | 0.013 |
| | **XGBoost** | 0.340 | 0.010 |
| | **Random Forest** | 0.425 | 0.014 |
| **RMSE** | **Our work** | 0.890 | 0.011 |
| | **SVM** | 2.219 | 0.028 |
| | **XGBoost** | 1.292 | 0.022 |
| | **Random Forest** | 1.306 | 0.032 |

*Note: For η, $R^2$ is evaluated in the $\log_{10}$-transformed space; MAE and RMSE are reported in the native linear space (cP). For ρ, all metrics are reported in $g/cm^3$. All four models are evaluated using the identical metric conventions to ensure fair comparison.*

As summarized in Table 2, Uni-Mix consistently outperforms all conventional baselines across $R^2$, MAE, and RMSE on both prediction tasks, with especially substantial reductions in absolute prediction errors for dynamic viscosity. This performance gap reflects not merely a difference in model capacity, but the fundamental representational limitations of fixed, predefined descriptors employed by traditional ML methods. Unlike density, viscosity is not an additive property that emerges from cooperative intermolecular interactions whose effects on flow resistance are highly sensitive to the precise three-dimensional arrangement of molecules. Conventional descriptors, however well engineered, reduce molecular structure to a fixed set of numerical features and in doing so inevitably discard the spatial relationships that govern how two distinct molecules interact at varying compositions. Our current framework circumvents this limitation by operating directly on 3D molecular conformations generated from SMILES strings. Its Transformer backbone, inheriting from Uni-Mol, fully connects all atoms regardless of bond connectivity and captures long-range interactions that are critical for hydrogen-bonding and polar systems but are missed by conventional descriptor-based methods. Moreover, the model is pre-trained on 209 million molecular conformations via self-supervised 3D coordinate recovery, acquiring a rich representation of geometry-property relationships that is transferable to mixture systems without requiring explicit parameterization of inter-component interactions. For reference, the MAE in $\log_{10}$(cP) given by our model is 0.035, surpassing the 0.043 reported in a recent GNN-based study on binary mixture viscosity[2]. This further highlights the benefit of leveraging 3D molecular pre-training in this task.

## Specific models for esters and hydrocarbons

To investigate whether specific fine-tuning confers predictive advantages over a broadly trained general model, specialized models for hydrocarbon and ester/anhydride binary mixtures are developed by extracting the corresponding subsets directly from the general dataset. As summarized in Table 3, these subsets are substantially smaller than the full training corpus. Each specialized model is fine-tuned independently on its own subset, whereas the general model is the aforementioned Uni-Mix model that is fine-tuned over the entire dataset.

**Table 3 | Dataset partitioning for dynamic viscosity and density prediction tasks across general, hydrocarbon-specific, and ester/anhydride-specific models.**

| Target | Dataset type | Total samples | Training set | Test set |
|---|---|---|---|---|
| **η** | General | 34,305 | 27,125 | 7,180 |
| | Hydrocarbons | 2,964 | 2,591 | 373 |
| | Esters & anhydrides | 991 | 835 | 156 |
| **ρ** | General | 28,816 | 23,576 | 5,240 |
| | Hydrocarbons | 3,897 | 3,290 | 607 |
| | Esters & anhydrides | 651 | 520 | 131 |

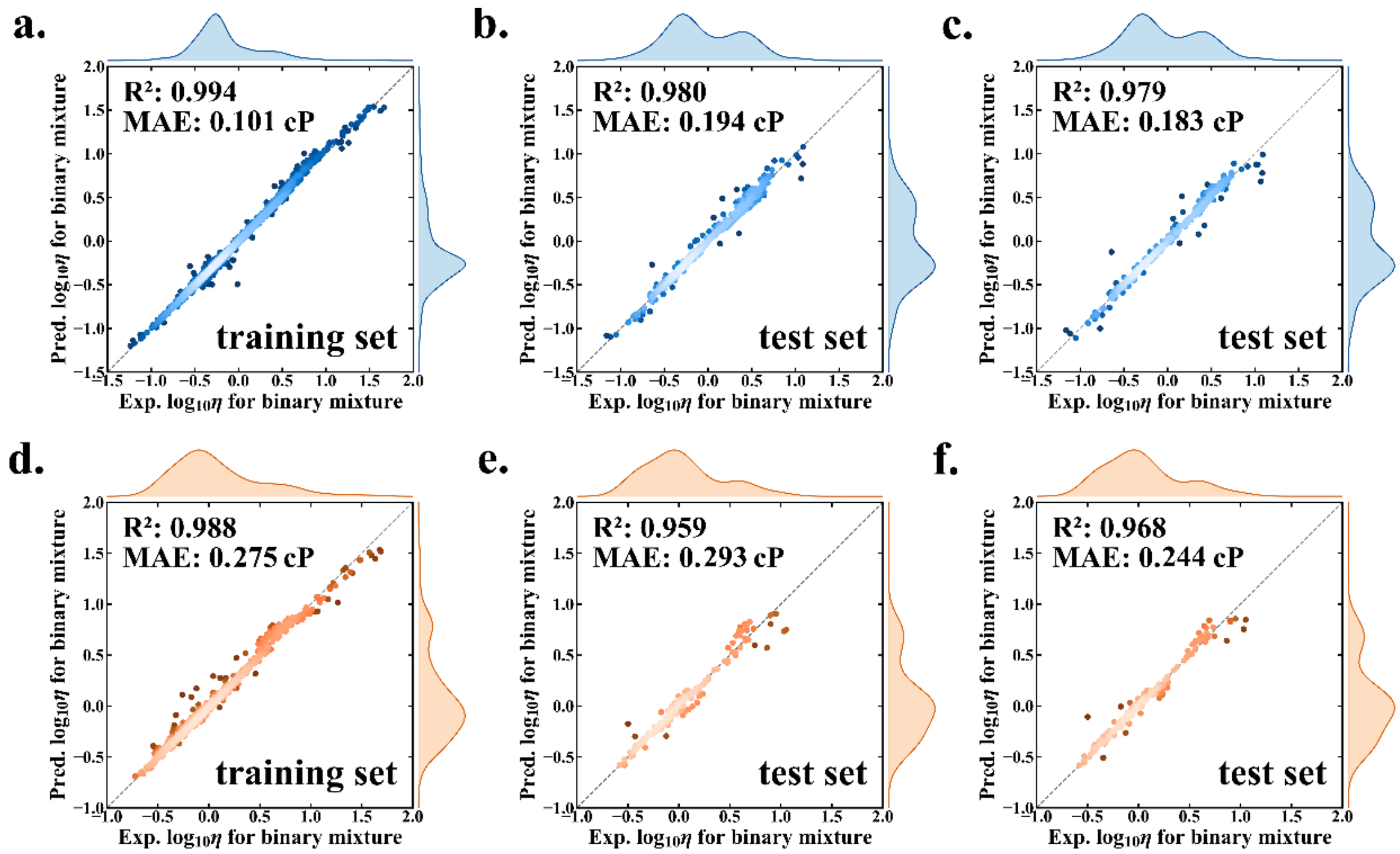

**Figure 3 | Performance of the Uni-Mix model in predicting dynamic viscosity for hydrocarbon and ester/anhydride binary systems.** ***a*** Hydrocarbon-specific model predictions on the hydrocarbon training set. ***b*** Hydrocarbon-specific model predictions on the hydrocarbon test set. ***c*** General model predictions on the hydrocarbon test set. ***d*** Ester/anhydride-specific model predictions on the ester/anhydride training set. ***e*** Ester/anhydride-specific model predictions on the ester/anhydride test set. ***f*** General model predictions on the ester/anhydride test set. Dashed lines represent ideal y = x correspondence; $R^2$ and MAE values are indicated in each panel.

To directly evaluate the impact of training scope on prediction accuracy, both the specialized and general models are assessed on the same test sets. As shown in Figure 3, for hydrocarbon systems, the general model achieves essentially equivalent performance to the hydrocarbon-specific model, with both yielding $R^2$ values of approximately 0.98, consistent with the relatively simple molecular structure and dispersion-force-dominated interactions of these systems[3,46]. For ester/anhydride systems, however, the general model outperforms the specialized counterpart, reducing MAE from 0.293 to 0.244 cP. The specialized model suffers from the limited size of the ester/anhydride training subset, which prevents adequate coverage of the structural diversity within this functional group family. The general model, by contrast, benefits from cross-category learning across a chemically diverse corpus, acquiring richer intermolecular representations that transfer effectively to ester/anhydride systems.

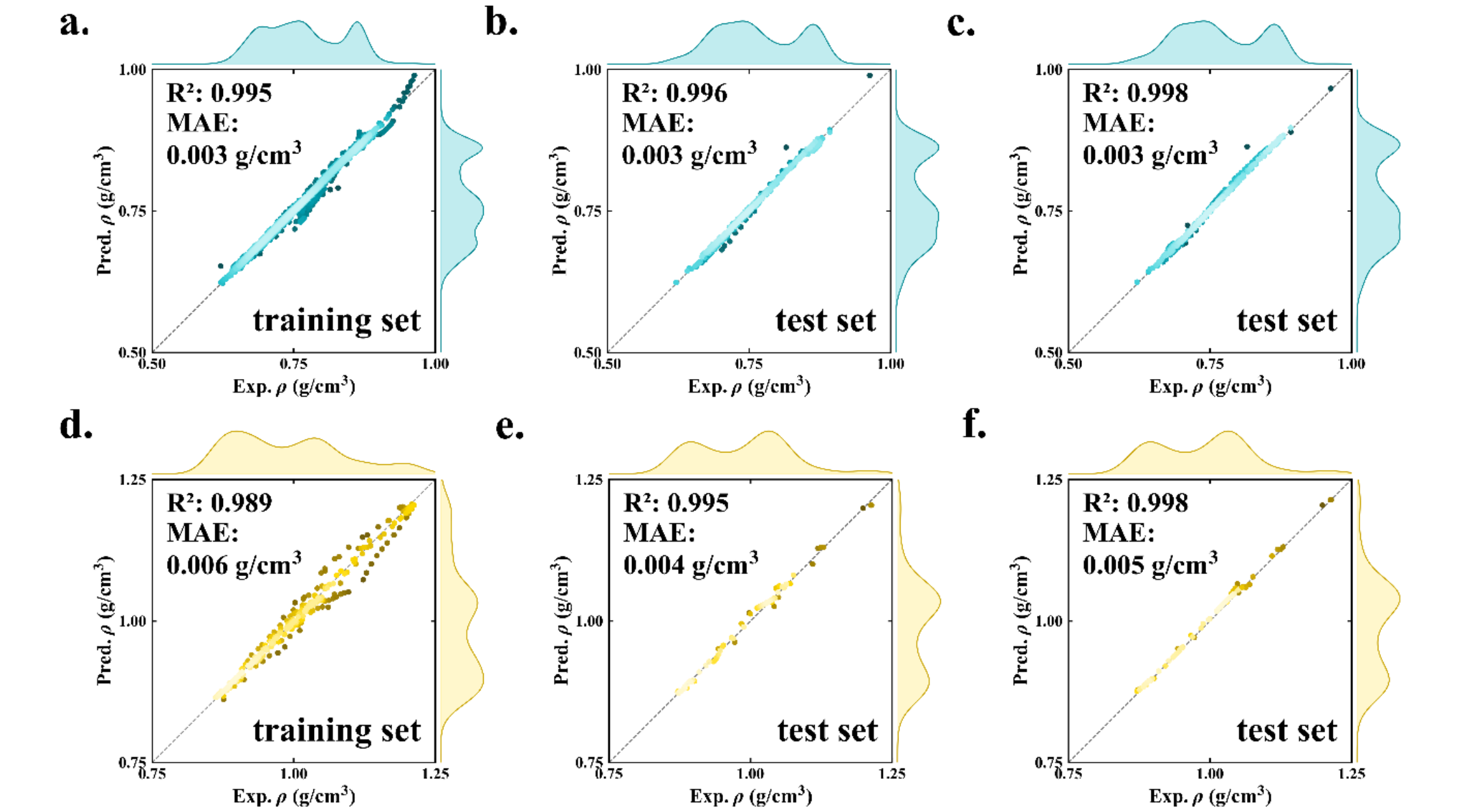

**Figure 4 | Performance of the Uni-Mix model in predicting density for hydrocarbon and ester/anhydride binary systems.** ***a*** Hydrocarbon-specific model predictions on the hydrocarbon training set. ***b*** Hydrocarbon-specific model predictions on the hydrocarbon test set. ***c*** General model predictions on the hydrocarbon test set. ***d*** Ester/anhydride-specific model predictions on the ester/anhydride training set. ***e*** Ester/anhydride-specific model predictions on the ester/anhydride test set. ***f*** General model predictions on the ester/anhydride test set. Dashed lines represent ideal y = x correspondence; $R^2$ and MAE values are indicated in each panel.

Density prediction results, shown in Figure 4, reinforce these findings. For hydrocarbon systems, the general model again achieves essentially equivalent accuracy to the hydrocarbon-specific model (Figures 4a–c), confirming that the structure-density relationship for hydrocarbons is sufficiently regular to be well captured without system-specific training. For ester/anhydride systems, the general model achieves a higher $R^2$ for ester/anhydride systems, despite a slight increase in MAE (Figures 4d–f), consistent with the viscosity results and further corroborating the generalization advantage of cross-category training for polar functional group systems.

Taken together, the specialized models primarily serve a diagnostic purpose. Their performance relative to the general model quantifies the latter's generalization capability across distinct chemical families. The robustness of the general model confirms its consistent accuracy across chemical families, while a specialized model may be preferred when computational efficiency is the primary constraint and the target chemical family is well defined and sufficiently represented in the training data.

## Experimental validation of model accuracy

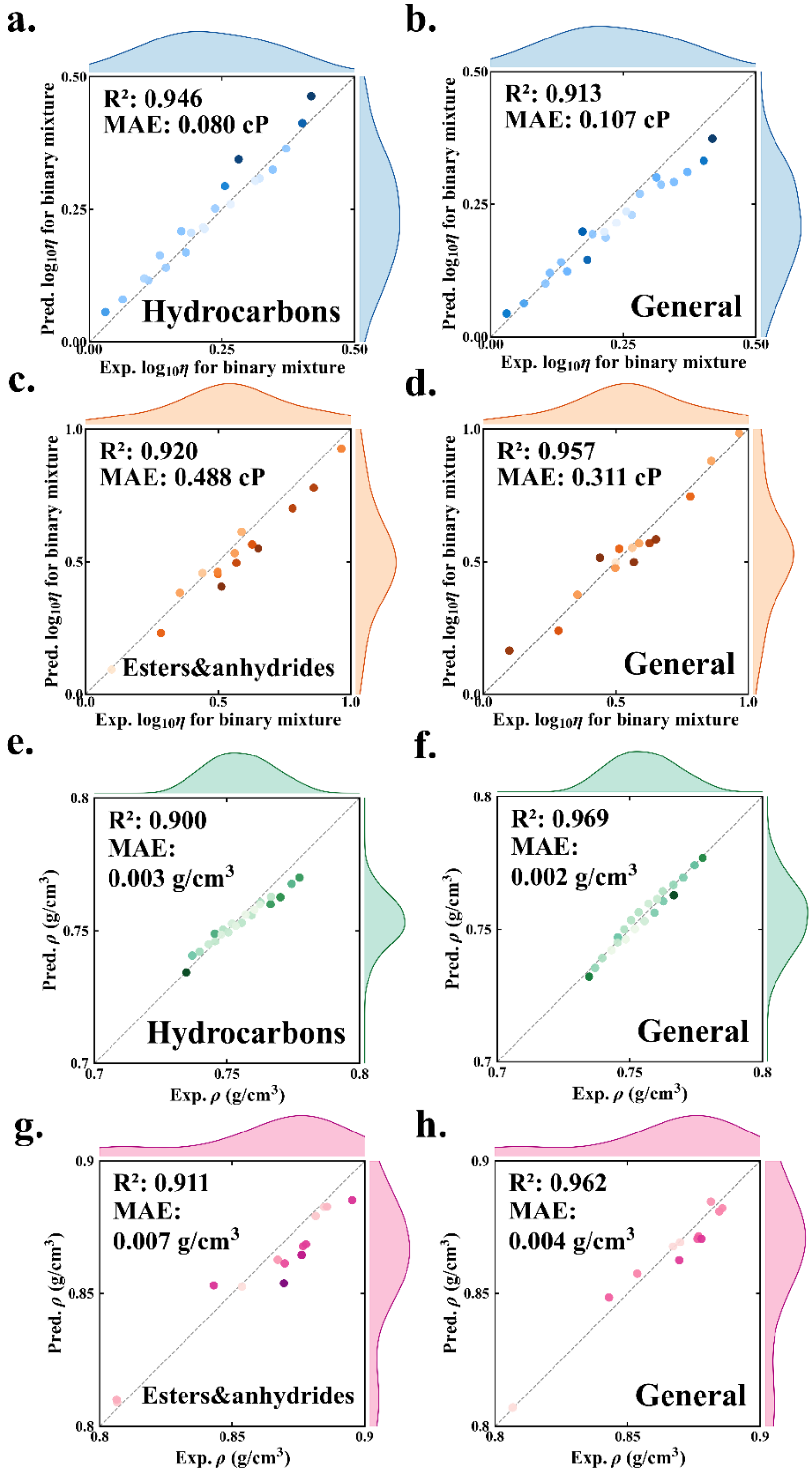

**Figure 5 | Comparison of model predictions with experimental data for independently prepared binary mixture samples.** ***a, b*** Comparison of experimental and predicted dynamic viscosity values for the

hydrocarbon dataset using the hydrocarbon-specific and general models. ***c, d*** Corresponding comparisons for the ester dataset. ***e–h*** Comparison of experimental and predicted density values for hydrocarbon and ester datasets. Error bars represent the repeatability ranges specified by ASTM D7042 for dynamic viscosity and ASTM D4052 for density. The general model provides superior or comparable accuracy across all tested systems.

To further validate the reliability of the developed models, experimental measurements are performed on independently prepared binary mixture samples. (See Table S6–S15 of the Supporting Information for the full set of validation data.) As shown in Figure 5, Uni-Mix predicted values are in good overall agreement with experimental measurements, confirming the practical reliability of the developed models. For dynamic viscosity on the hydrocarbon dataset, the specialized model achieves lower absolute error than the general model, indicating higher precision for hydrocarbon systems. It should be noted that such better performance of the hydrocarbon model reflects the difference between aggregate metrics across a diverse test set and performance on a specific, limited set of validation samples. For chemically regular families such as hydrocarbons, a specialized model can occasionally achieve finer quantitative agreement on small-scale experimental checks, whereas the general model offers more robust and transferable predictions across a broader chemical space. For the ester dataset, the general model markedly outperforms the ester-specific model, which is consistent with the conclusion we draw from the previous section. As detailed in Table S5 (with full predictions in Table S7 and Table S11), Uni-Mix, utilizing the same train/test splits and pair-grouped partitioning, demonstrates significantly better performance ($R^2$ = 0.913 for viscosity and 0.969 for density) than an XGBoost baseline ($R^2$ = 0.512 for viscosity and 0.392 for density).

To preliminarily explore the extensibility of the framework to multicomponent systems, prediction experiments are conducted on two ternary hydrocarbon mixtures not included in the training data, i.e., the mixture of n-dodecane, n-tetradecane, and 1-hexadecene as well as the mixture of n-tetradecane, 1-hexadecene, and 1-octadecene, comprising 12 data points at 298 and 313 K. As shown in Table 4, the model yields $R^2$ values of 0.928 and 0.944 for viscosity and density, respectively, indicating strong trend agreement between predicted and experimental values despite the small sample size (detailed in Table S14 and Table S15 of the Supporting Information), and suggesting promising transferability of the binary-trained model to ternary systems, although further validation on a broader range of mixtures is needed.

**Table 4 | Prediction performance of viscosity and density for ternary hydrocarbon mixtures.**

| Ternary system | Property | No. of combinations | $R^2$ | MAE |
|---|---|---|---|---|
| n-dodecane + n-tetradecane | $\log_{10}(\eta/\mathrm{cP})$ | 6 | 0.986 | 0.0388 |
| + 1-hexadecene | ρ (g/cm³) | 6 | 0.986 | 0.0160 |
| n-tetradecane + 1-hexadecene + 1-octadecene | $\log_{10}$(η/cP) | 6 | 0.980 | 0.1070 |
| | ρ (g/cm³) | 6 | 0.946 | 0.0263 |
| **Overall** | $\log_{10}$(η/cP) | 12 | 0.928 | 0.0729 |
| | ρ (g/cm³) | 12 | 0.944 | 0.0211 |

*Note: $R^2$ is reported as the squared Pearson correlation coefficient due to the small sample size. For viscosity, both $R^2$ and MAE are evaluated in $\log_{10}$ space; for density, in native linear space (g/cm3). All other tables use the standard $R^2$.*

## Impact of hydrogen bonding in viscosity prediction

Hydrogen-bonding systems, such as those containing alcohols or carboxylic acids, pose a greater challenge for viscosity prediction than non-hydrogen-bonding systems. Table 5 provides a statistical breakdown of prediction errors across different system types based on the complete dataset (training and test sets combined). It can be seen in Table 5 that notable performance differences are observed across system types. Non-hydrogen-bonding systems exhibit the highest $R^2$ and the lowest RMSE, as these systems are governed primarily by van der Waals and dispersion forces[44], which produce smooth, nearly linear viscosity-composition relationships that are straightforward for the model to capture. In contrast, alcohol-containing systems (which may contain considerable hydrogen bonds) yield an RMSE more than twice that of non-hydrogen-bonding systems, reflecting the strong nonlinearity introduced by dynamic, cooperative hydroxyl group networks whose topology varies sensitively with composition and temperature. Acid-containing systems present an intermediate case: although carboxyl groups form highly directional hydrogen bonds that give rise to stable dimer structures[45], the regularity of these interactions partially offsets the added complexity; nevertheless, prediction accuracy remains below that of non-hydrogen-bonding systems, and the limited number of acid-containing samples (2.22%) further constrains model learning. Overall, the presence of hydrogen bonding consistently increases prediction difficulty. Nevertheless, the model still achieves satisfactory accuracy across all system types ($R^2 \geq 0.803$ for alcohol-containing and acid-containing systems), demonstrating its robustness even for strongly interacting mixtures.

**Table 5 | Statistical analysis of viscosity prediction errors for different chemical system types classified by hydrogen bonding capability.**

| System type | Total samples | Data proportion | $R^2$ | RMSE (cP) |
|---|---|---|---|---|
| Alcohol-containing | 18,948 | 55.23% | 0.965 | 0.868 |
| Acid-containing | 762 | 2.22% | 0.803 | 0.873 |
| Non-H-bonding & other | 14,595 | 42.54% | 0.973 | 0.394 |

*Note: $R^2$ and RMSE are evaluated in the native linear space (cP) in this table to highlight the magnitude of absolute prediction errors across chemical system types.*

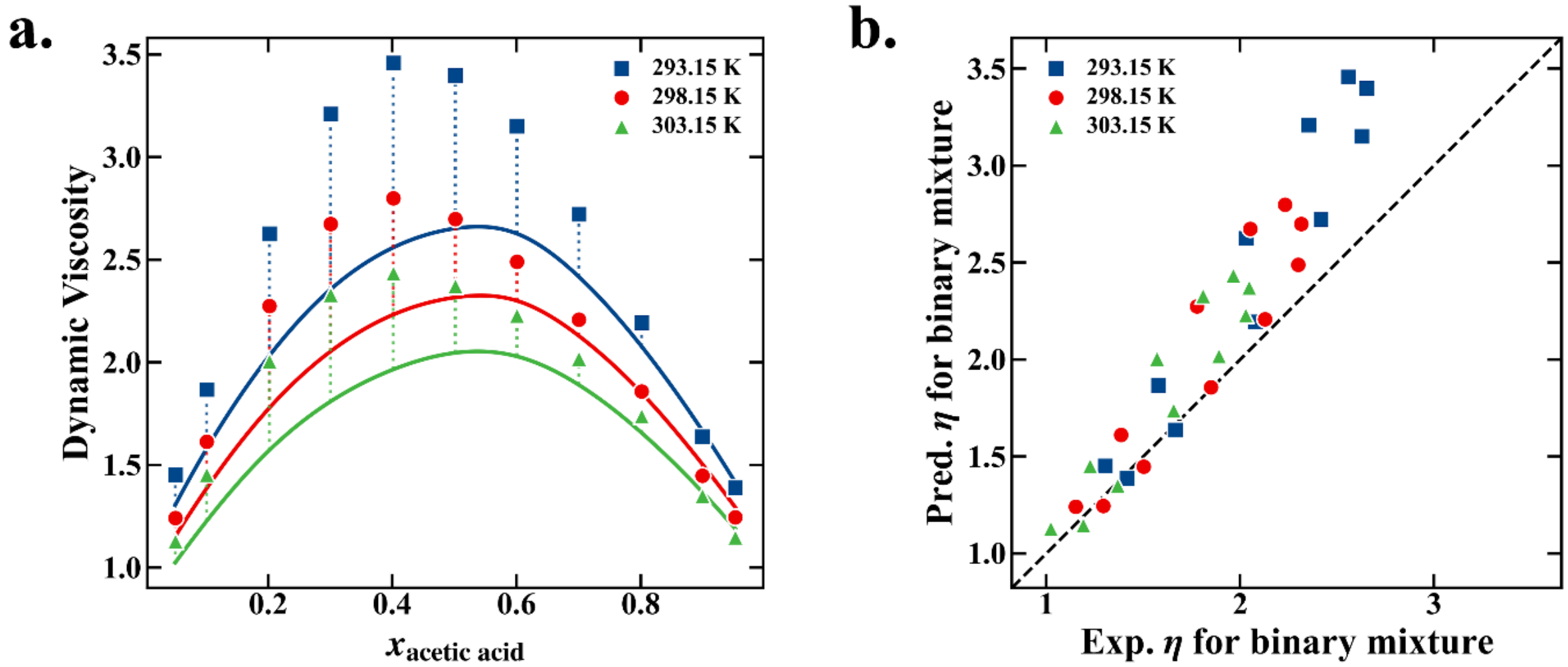

**Figure 6 | Non-monotonic viscosity behavior of the water-acetic acid binary mixture.** ***a*** Dynamic viscosity as a function of acetic acid mole fraction at three temperatures (293.15, 298.15, and 303.15 K). Solid lines represent experimental values; scatter points represent model predictions; vertical dashed lines indicate the deviation between predicted and experimental values at the same composition. ***b*** Parity plot comparing predicted and experimental viscosities of water-acetic mixture, with the dashed line indicating ideal agreement.

To further illustrate the model's capability in capturing the non-monotonic viscosity behavior of hydrogen-bonding systems, Figure 6 examines the water–acetic acid binary mixture, which exhibits pronounced non-monotonic behavior across 293.15–303.15 K. With increasing acetic acid mole fraction, the viscosity of the mixture first increases and then decreases, reaching a maximum near $x$ (acetic acid) ≈ 0.5~0.6. Notably, this maximum viscosity exceeds that of either pure component, indicating strong intermolecular interactions beyond ideal mixing[47]. The observed non-monotonic behavior likely arises from the formation of hydrogen-bonding networks between water and acetic acid, as well as the partial dissociation of acetic acid, which together enhance the resistance to flow. Despite this complexity, the Uni-Mix model successfully captures the non-monotonic trend, predicting the rise, peak, and subsequent decline of viscosity as a function of acetic acid content. The predicted values follow the experimental shape across the entire composition range, a capability that simple linear or logarithmic mixing rules completely lack. Some quantitative deviations remain, yet the model's ability to learn the non-ideal composition dependence from data demonstrates a clear advantage over empirical mixing rules in strongly interacting systems.

## Effect of hydrocarbon molecular type on viscosity-temperature behavior

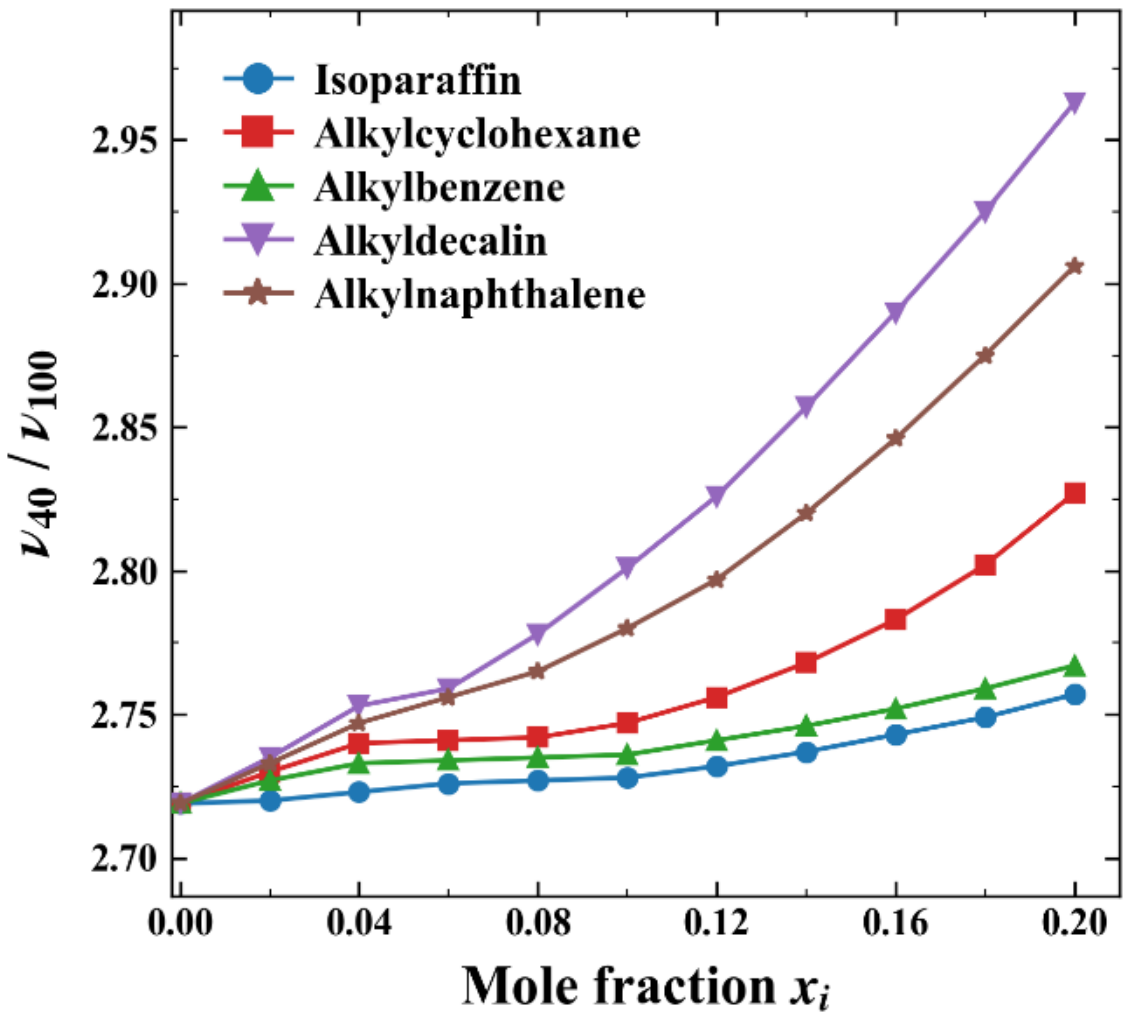

**Figure 7 | Viscosity-temperature sensitivity index $\nu_{40}/\nu_{100}$ as a function of component mole fraction $x_i$ for five binary hydrocarbon-PAO model systems with different molecular structures.** A higher $\nu_{40}/\nu_{100}$ ratio indicates greater temperature sensitivity and inferior viscosity-temperature performance. Five hydrocarbon types are compared, i.e., isoparaffin (H1), alkylcyclohexane (H2), alkylbenzene (H3), alkyldecalin (H4), and alkylnaphthalene (H5), while polyalphaolefin (PAO) is represented via a C30 saturated alkane model molecule. SMILES of these model compounds are given in Figure S1 and the predicted viscosity values in Table S16.

Having established the predictive accuracy of our model, we now explore how molecular structure affects viscosity-temperature behavior, the most critical property for lubricant design. Such analysis can guide the development of lubricants with improved performance across temperature. As shown in Figure 7, we examine five model hydrocarbon molecules with distinct structural characteristics, i.e., isoparaffin, alkylcyclohexane, alkylbenzene, alkyldecalin, and alkylnaphthalene. Each of these components is mixed with the same saturated C30 alkane model molecule representing polyalphaolefin (PAO) base oil. It can be seen that the ratio $\nu_{40}/\nu_{100}$ increases monotonically with mole fraction across all five systems, demonstrating that the addition of any of these hydrocarbon components consistently degrades the viscosity-temperature performance of the model PAO base oil.

The extent of degradation varies significantly with molecular structure among the five systems. Alkyldecalin and alkylnaphthalene show the most notable increases in $\nu_{40}/\nu_{100}$, reaching around 2.96 and 2.90, respectively, at a mole fraction of 0.20. This suggests that both naphthenic and aromatic multi-cyclic structures have the most detrimental impact on viscosity-temperature performance. Isoparaffin and alkylbenzene exhibit the least pronounced changes throughout the entire composition range, indicating that these two components have subtle influence to the viscosity index. Alkylcyclohexane demonstrates intermediate sensitivity, indicating the moderate effect of a single naphthenic ring. These findings highlight that even though hydrocarbon systems lack the distinct directional intermolecular interactions seen in hydrogen-bonding systems, the presence and nature of cyclic structures (especially fused rings) significantly influence the spatial arrangement and flow resistance within the model PAO matrix. Specifically, fused-ring and naphthenic groups (instead of single aromatic rings) cause the most disruption, aligning with previous computational research showing a negative correlation between increasing ring count and viscosity index[46] and further demonstrating the reliability of our predictions.

# Conclusion

In summary, this work demonstrates the potential of Uni-Mix, a 3D molecular representation learning framework, for predicting the viscosity and density of organic mixtures. By extending the model input interface to incorporate mixture-level descriptors and performing property-specific fine-tuning, structure-property mappings have been established that achieve $R^2$ values of 0.973 and 0.996 on the test sets for dynamic viscosity and density prediction, respectively, substantially outperforming conventional machine learning baselines. Such framework can be straightforwardly extended to other properties of organic mixtures, provided sufficient experimental data are available.

Systematic comparison between general and specialized models reveals that cross-category training on chemically diverse datasets achieves comparable accuracy to specialized models for structurally regular hydrocarbon systems, while conferring a marked generalization advantage for polar functional group systems where limited training data restrict specialized model performance. The model accurately captures nonlinear viscosity-composition relationships, temperature-dependent behavior, and the asymmetric influence of hydrogen bonding on prediction accuracy. Experimental validation on independently prepared samples confirms the practical reliability of the predictions. Moreover, we validate the framework on ternary hydrocarbon mixtures, demonstrating its direct applicability to multicomponent systems with promising predictive performance. Using the model, we further analyze how different hydrocarbon molecular structures—branching, cycloalkane rings, and aromatic rings—affect viscosity-temperature behavior, providing quantitative guidance for designing lubricants with improved viscosity-temperature performance.

Altogether, the demonstrated framework combines 3D molecular pre-training with mixture-level fine-tuning, providing a foundation for extension to higher-order mixtures and other thermophysical properties, and paves the way for the rational and efficient design of functional fluid formulations for energy-saving applications.

# Data and code availability

The fine-tuning datasets compiled from the Landolt–Börnstein database are described in the Methods section (with the corresponding references). Complete fine-tuning datasets can be obtained upon reasonable request by contacting the corresponding authors. Experimental validated results are provided for reproducibility in Table S6–S15 of the Supporting Information.

# Supporting Information Available

Detailed data pre-processing and split strategy; hyper-parameters for all fine-tuned models (Tables S1–S4); comparison of XGBoost versus Uni-Mix on independently measured experimental values (Table S5); experimental validation data for viscosity and density of binary hydrocarbon and ester/anhydride mixtures (Table S6–S13); prediction results for ternary hydrocarbon mixtures (Table S14–S15); molecular structures of five representative hydrocarbons and the PAO model molecule (Figure S1); predicted dynamic viscosity values for pure PAO and five binary PAO mixtures (Table S16).

# Acknowledgement

This work was supported by research grants from the National Natural Science Foundation of China (grant number 22573124), China Petroleum & Chemical Corp (funding number 124014), and Beijing Municipal Science & Technology Commission (grant number Z251100007525011).

# Competing Interests

The authors declare no conflict of interest.

# Supplementary Information for: 3D molecular representation learning for organic mixtures: viscosity and density prediction

Haicheng Qu[1ǂ], Yanyi Su[2,3,ǂ], Ning Wang[1], Shangqian Chen[3], Zhifeng Gao[3*], Jun Cheng[2,4,5*], Qi Ou[1*]

[1]Research Institute of Petroleum Processing, Sinopec Corporation, 18th Xueyuan Road, Haidian District, Beijing, China

[2]State Key Laboratory of Physical Chemistry of Solid Surface, College of Chemistry and Chemical Engineering, Xiamen University, Xiamen, China

[3]DP Technology, Beijing, China

[4]Laboratory of AI for Electrochemistry (AI4EC), Tan Kah Kee Innovation Laboratory (IKKEM), Xiamen, China

[5]Institute of Artificial Intelligence, Xiamen University, Xiamen, China

*Email: gaozf@dp.tech, chengjun@xmu.edu.cn, ouqi.ripp@sinopec.com

ǂThese authors contributed equally to this work.

# Data pre-processing and split strategy

The experimental mixture property data used for fine-tuning were compiled from the Landolt–Börnstein database (Refs. 41-42 of the main text and Refs 1-2 in this SI); the independently measured validation data are provided in Table S6–S15. The dynamic viscosity η was transformed to $\log_{10}\eta$ prior to training to account for the three-order-of-magnitude span of the data (0.05–60.8 cP), which equalizes sample contributions on a relative-error basis and embeds the Arrhenius-type temperature dependence as a physically meaningful prior; the density ρ was used without transformation. For viscosity, $R^2$ is computed in $\log_{10}$ space while MAE and RMSE are reported in the native linear space (cP) after back-transformation; for density, all metrics are in native linear space ($g/cm^3$). Unless otherwise noted, $R^2$ denotes the standard coefficient of determination. The train/test split is grouped by the SMILES pair: all records sharing the same molecular pair—across all mole fractions and temperatures—are assigned as a unit, ensuring that no mixture pair in the test set appears during training. The pre-trained Uni-Mix model is accessible via `https://funmg.dp.tech/uni-mix`, while the original training and test data sets of all aforementioned properties (including the corresponding PM6-optimized 3D coordinates) are compressed in the supplementary file `data.zip` with a csv format.

# Nomenclature

The following symbols and abbreviations are used throughout this Supporting Information and the main text.

| Symbol | Definition |
|---|---|
| η | Dynamic viscosity (cP) |
| ρ | Density ($g/cm^3$) |
| ν | Kinematic viscosity(cSt) |
| T | Absolute temperature (K) |
| xi | Mole fraction of component i |
| SMILES | Simplified Molecular Input Line Entry System |

# Model Training Parameters

Hyper-parameters for each fine-tuning process of the random splitting test are given as follows:

**Table S1. Hyper-parameters for fine-tuning.**

| Parameter | η General | η Hydro | η Ester | ρ General | ρ Hydro | ρ Ester |
|---|---|---|---|---|---|---|
| task | regression | regression | regression | regression | regression | regression |

| data_type | molecule | molecule | molecule | molecule | molecule | molecule |
|---|---|---|---|---|---|---|
| split | mixture | mixture | mixture | mixture | mixture | mixture |
| epochs | 1000 | 1000 | 1000 | 1000 | 1000 | 1000 |
| batch_size | 32 | 32 | 32 | 32 | 32 | 32 |
| learning_rate | 4e-5 | 4e-5 | 4e-5 | 4e-5 | 4e-5 | 4e-5 |
| early_stopping | 30 | 30 | 30 | 30 | 30 | 30 |
| metrics | r2 | r2 | r2 | r2 | r2 | r2 |
| target_normalize | log | log | log | log | log | log |
| model_name | unimolv1 | unimolv1 | unimolv1 | unimolv1 | unimolv1 | unimolv1 |
| chemix_embed_dim | 256 | 256 | 256 | 256 | 256 | 256 |
| chemix_attn_layers | 4 | 4 | 4 | 4 | 4 | 4 |
| chemix_attn_heads | 8 | 8 | 8 | 8 | 8 | 8 |
| loss_key | nll | nll | nll | nll | nll | nll |

Three baselines (Random Forest, SVR with RBF kernel, and XGBoost) were trained on the same train/test splits as the Uni-Mol general model. Hyper-parameters were selected via grid search with 5-fold cross-validation on the training set. Selected hyper-parameters are listed in Tables S2–S4; all other parameters use the respective library defaults. Viscosity was trained on $\log10(\eta)$, density on $\rho$ directly.

**Table S2. Selected hyper-parameters of the Random Forest baseline.**

| Parameter | η (general) | ρ (general) |
|---|---|---|
| n_estimators | 1000 | 600 |
| max_depth | 20 | 20 |
| max_features | 0.7 | 0.5 |
| min_samples_split | 2 | 5 |
| min_samples_leaf | 1 | 1 |

**Table S3. Selected hyper-parameters of the SVR baseline (RBF kernel).**

| Parameter | η (general) | ρ (general) |
|---|---|---|

| C | 100 | 100 |
|---|---|---|
| epsilon | 0.01 | 0.1 |
| gamma | 0.001 | 0.01 |

**Table S4. Selected hyper-parameters of the XGBoost baseline.**

| Parameter | η (general) | ρ (general) |
|---|---|---|
| learning_rate | 0.05 | 0.05 |
| n_estimators | 3149 | 3024 |
| max_depth | 6 | 6 |
| min_child_weight | 1 | 3 |
| subsample | 1.0 | 0.7 |
| colsample_bytree | 0.7 | 0.85 |
| reg_alpha | 0.1 | 0.0 |
| reg_lambda | 0.1 | 0.1 |

**Table S5. Comparison of XGBoost and Uni-Mix predictions against independently measured experimental values for hydrocarbon binary mixtures.** Both models were fine-tuned on the hydrocarbon subset of the Landolt–Börnstein corpus, utilizing identical train/test splits and pair-grouped partitioning. The XGBoost baseline employs a 161-dimensional hand-crafted feature vector for each sample, which includes 16 RDKit 2D descriptors for each component, 20 statistics derived from SMILES strings, mole-fraction-weighted and element-wise-difference versions of these vectors, and nine auxiliary scalars that encode composition and temperature. In contrast, Uni-Mix utilizes a 512-dimensional pre-trained Uni-Mol 3D embedding for each component, concatenated with a compact 15-dimensional block of RDKit descriptors.

| Model | Viscosity | | Density | |
|---|---|---|---|---|
| | $R^2$ ($\log_{10}\eta$) | MAE (cP) | $R^2$ ($\rho$) | MAE (g/cm$^3$) |
| XGBoost | 0.512 | 0.305 | 0.392 | 0.007 |
| **Uni-Mix** | **0.913** | **0.107** | **0.969** | **0.002** |

*$R^2$ for viscosity is computed in $log_{10}$ space; MAE is reported in the native linear space after back-transformation. Uni-Mix predictions in Table S7 and Table S11.*

# Experimental Table

**Table S6 Comparison of experimental and predicted dynamic viscosity values for the hydrocarbon binary mixture dataset using the hydrocarbon-specific model.**

| SMILES_1 | SMILES_2 | x1 | x2 | *T*(K) | $\eta_{exp}$ (cP) | $\eta_{pred}$ (cP) |
|---|---|---|---|---|---|---|
| CCCCCCCCCCCCCC | CCCCCCCCCCCCCCC=C | 1 | 0 | 298 | 2.0532 | 2.0139 |
| CCCCCCCCCCCCCC | CCCCCCCCCCCCCCC=C | 0.8190 | 0.1810 | 298 | 2.0964 | 2.0349 |
| CCCCCCCCCCCCCC | CCCCCCCCCCCCCCC=C | 0.6292 | 0.3708 | 298 | 2.2179 | 2.1126 |
| CCCCCCCCCCCCCC | CCCCCCCCCCCCCCC=C | 0.4299 | 0.5701 | 298 | 2.3494 | 2.3144 |
| CCCCCCCCCCCCCC | CCCCCCCCCCCCCCC=C | 0.2205 | 0.7795 | 298 | 2.5234 | 2.5833 |
| CCCCCCCCCCCCCC | CCCCCCCCCCCCCCC=C | 0 | 1 | 298 | 2.6210 | 2.9059 |
| CCCCCCCCCCCCCC | CCCCCCCCCCCC | 1 | 0 | 298 | 2.0532 | 2.0139 |
| CCCCCCCCCCCCCC | CCCCCCCCCCCC | 0.7745 | 0.2255 | 298 | 1.8472 | 1.8178 |
| CCCCCCCCCCCCCC | CCCCCCCCCCCC | 0.5629 | 0.4371 | 298 | 1.6483 | 1.6315 |
| CCCCCCCCCCCCCC | CCCCCCCCCCCC | 0.3640 | 0.6360 | 298 | 1.5207 | 1.4750 |
| CCCCCCCCCCCCCC | CCCCCCCCCCCC | 0.1767 | 0.8233 | 298 | 1.3945 | 1.3795 |
| CCCCCCCCCCCCCC | CCCCCCCCCCCC | 0 | 1 | 298 | 1.2922 | 1.3056 |
| CCCCCCCCCCCCCC | CCCCCCCCCCCCCCC=C | 1 | 0 | 313 | 1.4895 | 1.6155 |
| CCCCCCCCCCCCCC | CCCCCCCCCCCCCCC=C | 0.8190 | 0.1810 | 313 | 1.5568 | 1.6053 |
| CCCCCCCCCCCCCC | CCCCCCCCCCCCCCC=C | 0.6292 | 0.3708 | 313 | 1.6381 | 1.6466 |
| CCCCCCCCCCCCCC | CCCCCCCCCCCCCCC=C | 0.4299 | 0.5701 | 313 | 1.7253 | 1.7844 |
| CCCCCCCCCCCCCC | CCCCCCCCCCCCCCC=C | 0.2205 | 0.7795 | 313 | 1.8026 | 1.9670 |
| CCCCCCCCCCCCCC | CCCCCCCCCCCCCCC=C | 0 | 1 | 313 | 1.9126 | 2.2082 |
| CCCCCCCCCCCCCC | CCCCCCCCCCCC | 1 | 0 | 313 | 1.4895 | 1.6155 |
| CCCCCCCCCCCCCC | CCCCCCCCCCCC | 0.7745 | 0.2255 | 313 | 1.3588 | 1.4569 |
| CCCCCCCCCCCCCC | CCCCCCCCCCCC | 0.5629 | 0.4371 | 313 | 1.2678 | 1.3170 |
| CCCCCCCCCCCCCC | CCCCCCCCCCCC | 0.3640 | 0.6360 | 313 | 1.1561 | 1.2025 |
| CCCCCCCCCCCCCC | CCCCCCCCCCCC | 0.1767 | 0.8233 | 313 | 1.0713 | 1.1379 |

**Table S7 Comparison of experimental and predicted dynamic viscosity values for the hydrocarbon binary mixture dataset using the general model.**

| SMILES_1 | SMILES_2 | x1 | x2 | $T$(K) | $\eta_{exp}$ (cP) | $\eta_{pred}$ (cP) |
|---|---|---|---|---|---|---|
| CCCCCCCCCCCCCC | CCCCCCCCCCCCCCC=C | 1 | 0 | 298 | 2.0532 | 1.9981 |
| CCCCCCCCCCCCCC | CCCCCCCCCCCCCCC=C | 0.8190 | 0.1810 | 298 | 2.0964 | 1.9366 |
| CCCCCCCCCCCCCC | CCCCCCCCCCCCCCC=C | 0.6292 | 0.3708 | 298 | 2.2179 | 1.9589 |
| CCCCCCCCCCCCCC | CCCCCCCCCCCCCCC=C | 0.4299 | 0.5701 | 298 | 2.3494 | 2.0448 |
| CCCCCCCCCCCCCC | CCCCCCCCCCCCCCC=C | 0.2205 | 0.7795 | 298 | 2.5234 | 2.1459 |
| CCCCCCCCCCCCCC | CCCCCCCCCCCCCCC=C | 0 | 1 | 298 | 2.6210 | 2.3641 |
| CCCCCCCCCCCCCC | CCCCCCCCCCCC | 1 | 0 | 298 | 2.0532 | 1.9981 |
| CCCCCCCCCCCCCC | CCCCCCCCCCCC | 0.7745 | 0.2255 | 298 | 1.8472 | 1.6979 |
| CCCCCCCCCCCCCC | CCCCCCCCCCCC | 0.5629 | 0.4371 | 298 | 1.6483 | 1.5376 |
| CCCCCCCCCCCCCC | CCCCCCCCCCCC | 0.3640 | 0.6360 | 298 | 1.5207 | 1.3975 |
| CCCCCCCCCCCCCC | CCCCCCCCCCCC | 0.1767 | 0.8233 | 298 | 1.3945 | 1.3273 |
| CCCCCCCCCCCCCC | CCCCCCCCCCCC | 0 | 1 | 298 | 1.2922 | 1.3184 |
| CCCCCCCCCCCCCC | CCCCCCCCCCCCCCC=C | 1 | 0 | 313 | 1.4895 | 1.5770 |
| CCCCCCCCCCCCCC | CCCCCCCCCCCCCCC=C | 0.8190 | 0.1810 | 313 | 1.5568 | 1.5600 |
| CCCCCCCCCCCCCC | CCCCCCCCCCCCCCC=C | 0.6292 | 0.3708 | 313 | 1.6381 | 1.5750 |
| CCCCCCCCCCCCCC | CCCCCCCCCCCCCCC=C | 0.4299 | 0.5701 | 313 | 1.7253 | 1.6403 |
| CCCCCCCCCCCCCC | CCCCCCCCCCCCCCC=C | 0.2205 | 0.7795 | 313 | 1.8026 | 1.7228 |
| CCCCCCCCCCCCCC | CCCCCCCCCCCCCCC=C | 0 | 1 | 313 | 1.9126 | 1.8577 |
| CCCCCCCCCCCCCC | CCCCCCCCCCCC | 1 | 0 | 313 | 1.4895 | 1.5770 |
| CCCCCCCCCCCCCC | CCCCCCCCCCCC | 0.7745 | 0.2255 | 313 | 1.3588 | 1.3823 |
| CCCCCCCCCCCCCC | CCCCCCCCCCCC | 0.5629 | 0.4371 | 313 | 1.2678 | 1.2597 |
| CCCCCCCCCCCCCC | CCCCCCCCCCCC | 0.3640 | 0.6360 | 313 | 1.1561 | 1.1560 |
| CCCCCCCCCCCCCC | CCCCCCCCCCCC | 0.1767 | 0.8233 | 313 | 1.0713 | 1.1057 |

**Table S8 Comparison of experimental and predicted dynamic viscosity values for the ester and anhydride binary mixture dataset using the ester & anhydride-specific model.**

| SMILES_1 | SMILES_2 | x1 | x2 | $T$(K) | $\eta_{exp}$ (cP) | $\eta_{pred}$ (cP) |
|---|---|---|---|---|---|---|
| CCCCCCCCCCCCCOC(=O)C(=C)C | CC(C)CCCCC(=O)OCCCCCCCCCC | 1 | 0 | 298 | 3.8743 | 4.0907 |
| CCCCCCCCCCCCCOC(=O)C(=C)C | CC(C)CCCCC(=O)OCCCCCCCCCC | 0.8173 | 0.1827 | 298 | 3.6517 | 3.4065 |
| CCCCCCCCCCCCCOC(=O)C(=C)C | CC(C)CCCCC(=O)OCCCCCCCCCC | 0.6265 | 0.3735 | 298 | 3.7045 | 3.1320 |
| CCCCCCCCCCCCCOC(=O)C(=C)C | CC(C)CCCCC(=O)OCCCCCCCCCC | 0.4271 | 0.5729 | 298 | 4.2400 | 3.6723 |
| CCCCCCCCCCCCCOC(=O)C(=C)C | CC(C)CCCCC(=O)OCCCCCCCCCC | 0.2185 | 0.7815 | 298 | 6.0335 | 5.0254 |
| CCCCCCCCCCCCCOC(=O)C(=C)C | CC(C)CCCCC(=O)OCCCCCCCCCC | 0 | 1 | 298 | 9.2265 | 8.4437 |
| CCCCCCCCCCCCCOC(=O)C(=C)C | CC(C)CCCCC(=O)OCCCCCCCCCC | 1 | 0 | 313 | 2.7581 | 2.8618 |
| CCCCCCCCCCCCCOC(=O)C(=C)C | CC(C)CCCCC(=O)OCCCCCCCCCC | 0.8173 | 0.1827 | 313 | 3.1459 | 2.8908 |
| CCCCCCCCCCCCCOC(=O)C(=C)C | CC(C)CCCCC(=O)OCCCCCCCCCC | 0.6265 | 0.3735 | 313 | 3.1543 | 2.8410 |
| CCCCCCCCCCCCCOC(=O)C(=C)C | CC(C)CCCCC(=O)OCCCCCCCCCC | 0.4271 | 0.5729 | 313 | 3.2535 | 2.5493 |
| CCCCCCCCCCCCCOC(=O)C(=C)C | CC(C)CCCCC(=O)OCCCCCCCCCC | 0.2185 | 0.7815 | 313 | 4.4772 | 3.5458 |
| CCCCCCCCCCCCCOC(=O)C(=C)C | CC(C)CCCCC(=O)OCCCCCCCCCC | 0 | 1 | 313 | 7.2516 | 6.0112 |
| CCCCC(CC)C(=O)OCC(COC(=O)C(CC)CCCC)(COC(=O | CCCCCCOC(=O)C(=C)C | 0.1504 | 0.8496 | 298 | 2.2635 | 2.4157 |
| )C(CC)CCCC)COC(=O)C(CC)CCCC | | | | | | |
| CCCCC(CC)C(=O)OCC(COC(=O)C(CC)CCCC)(COC(=O)C(CC)CCCC)COC(=O)C(CC)CCCC | CCCCCCOC(=O)C(=C)C | 0.0623 | 0.9377 | 298 | 1.9244 | 1.7044 |
| CCCCC(CC)C(=O)OCC(COC(=O)C(CC)CCCC)(COC(=O)C(CC)CCCC)COC(=O)C(CC)CCCC | CCCCCCOC(=O)C(=C)C | 0 | 1 | 298 | 1.2525 | 1.2412 |

**Table S9 Comparison of experimental and predicted dynamic viscosity values for the ester and anhydride binary mixture dataset using the general model.**

| SMILES_1 | SMILES_2 | x1 | x2 | $T$(K) | $\eta_{exp}$ (cP) | $\eta_{pred}$ (cP) |
|---|---|---|---|---|---|---|
| CCCCCCCCCCCCOC(=O)C(=C)C | CC(C)CCCCC(=O)OCCCCCCCCC | 1 | 0 | 298 | 3.8743 | 3.7072 |
| CCCCCCCCCCCCOC(=O)C(=C)C | CC(C)CCCCC(=O)OCCCCCCCCC | 0.8173 | 0.1827 | 298 | 3.6517 | 3.5648 |
| CCCCCCCCCCCCOC(=O)C(=C)C | CC(C)CCCCC(=O)OCCCCCCCCC | 0.6265 | 0.3735 | 298 | 3.7045 | 3.1510 |
| CCCCCCCCCCCCOC(=O)C(=C)C | CC(C)CCCCC(=O)OCCCCCCCCC | 0.4271 | 0.5729 | 298 | 4.2400 | 3.7104 |
| CCCCCCCCCCCCOC(=O)C(=C)C | CC(C)CCCCC(=O)OCCCCCCCCC | 0.2185 | 0.7815 | 298 | 6.0335 | 5.5587 |
| CCCCCCCCCCCCOC(=O)C(=C)C | CC(C)CCCCC(=O)OCCCCCCCCC | 0 | 1 | 298 | 9.2265 | 9.6475 |
| CCCCCCCCCCCCOC(=O)C(=C)C | CC(C)CCCCC(=O)OCCCCCCCCC | 1 | 0 | 313 | 2.7581 | 3.2753 |
| CCCCCCCCCCCCOC(=O)C(=C)C | CC(C)CCCCC(=O)OCCCCCCCCC | 0.8173 | 0.1827 | 313 | 3.1459 | 2.9906 |
| CCCCCCCCCCCCCOC(=O)C(=C)C | CC(C)CCCCC(=O)OCCCCCCCCCC | 0.6265 | 0.3735 | 313 | 3.1543 | 3.1402 |
| CCCCCCCCCCCCCOC(=O)C(=C)C | CC(C)CCCCC(=O)OCCCCCCCCCC | 0.4271 | 0.5729 | 313 | 3.2535 | 3.5347 |
| CCCCCCCCCCCCCOC(=O)C(=C)C | CC(C)CCCCC(=O)OCCCCCCCCCC | 0.2185 | 0.7815 | 313 | 4.4772 | 3.8345 |
| CCCCCCCCCCCCCOC(=O)C(=C)C | CC(C)CCCCC(=O)OCCCCCCCCCC | 0 | 1 | 313 | 7.2516 | 7.5716 |
| CCCCC(CC)C(=O)OCC(COC(=O)C(CC)CCCC)(COC(=O)C(CC)CCCC)COC(=O)C(CC)CCCC | CCCCCCOC(=O)C(=C)C | 0.1504 | 0.8496 | 298 | 2.2635 | 2.3758 |
| CCCCC(CC)C(=O)OCC(COC(=O)C(CC)CCCC)(COC(=O)C(CC)CCCC)COC(=O)C(CC)CCCC | CCCCCCOC(=O)C(=C)C | 0.0623 | 0.9377 | 298 | 1.9244 | 1.7380 |
| CCCCC(CC)C(=O)OCC(COC(=O)C(CC)CCCC)(COC(=O)C(CC)CCCC)COC(=O)C(CC)CCCC | CCCCCCOC(=O)C(=C)C | 0 | 1 | 298 | 1.2525 | 1.4587 |

**Table S10 Comparison of experimental and predicted density values for the hydrocarbon binary mixture dataset using the hydrocarbon-specific model.**

| SMILES_1 | SMILES_2 | x1 | x2 | *T*(K) | $\rho_{exp}$ (g/cm$^3$) | $\rho_{pred}$ (g/cm$^3$) |
|---|---|---|---|---|---|---|
| CCCCCCCCCCCCCCC | CCCCCCCCCCCCCCC=C | 1 | 0 | 298 | 0.7603 | 0.7579 |
| CCCCCCCCCCCCCCC | CCCCCCCCCCCCCCC=C | 0.8190 | 0.1810 | 298 | 0.7625 | 0.7599 |
| CCCCCCCCCCCCCCC | CCCCCCCCCCCCCCC=C | 0.6292 | 0.3708 | 298 | 0.7665 | 0.7600 |
| CCCCCCCCCCCCCCC | CCCCCCCCCCCCCCC=C | 0.4299 | 0.5701 | 298 | 0.7702 | 0.7627 |
| CCCCCCCCCCCCCCC | CCCCCCCCCCCCCCC=C | 0.2205 | 0.7795 | 298 | 0.7744 | 0.7677 |
| CCCCCCCCCCCCCCC | CCCCCCCCCCCCCCC=C | 0 | 1 | 298 | 0.7775 | 0.7700 |

| CCCCCCCCCCCCCC | CCCCCCCCCCCC | 1 | 0 | 298 | 0.7603 | 0.7579 |
|---|---|---|---|---|---|---|
| CCCCCCCCCCCCCC | CCCCCCCCCCCC | 0.7745 | 0.2255 | 298 | 0.7571 | 0.7561 |
| CCCCCCCCCCCCCC | CCCCCCCCCCCC | 0.5629 | 0.4371 | 298 | 0.7534 | 0.7519 |
| CCCCCCCCCCCCCC | CCCCCCCCCCCC | 0.3640 | 0.6360 | 298 | 0.7506 | 0.7495 |
| CCCCCCCCCCCCCC | CCCCCCCCCCCC | 0.1767 | 0.8233 | 298 | 0.7479 | 0.7486 |
| CCCCCCCCCCCCCC | CCCCCCCCCCCC | 0 | 1 | 298 | 0.7455 | 0.7459 |
| CCCCCCCCCCCCCC | CCCCCCCCCCCCCCC=C | 1 | 0 | 313 | 0.7485 | 0.7505 |
| CCCCCCCCCCCCCC | CCCCCCCCCCCCCCC=C | 0.8190 | 0.1810 | 313 | 0.7520 | 0.7527 |
| CCCCCCCCCCCCCC | CCCCCCCCCCCCCCC=C | 0.6292 | 0.3708 | 313 | 0.7556 | 0.7530 |
| CCCCCCCCCCCCCC | CCCCCCCCCCCCCCC=C | 0.4299 | 0.5701 | 313 | 0.7593 | 0.7559 |
| CCCCCCCCCCCCCC | CCCCCCCCCCCCCCC=C | 0.2205 | 0.7795 | 313 | 0.7627 | 0.7608 |
| CCCCCCCCCCCCCC | CCCCCCCCCCCCCCC=C | 0 | 1 | 313 | 0.7667 | 0.7628 |
| CCCCCCCCCCCCCC | CCCCCCCCCCCC | 1 | 0 | 313 | 0.7485 | 0.7505 |
| CCCCCCCCCCCCCC | CCCCCCCCCCCC | 0.7745 | 0.2255 | 313 | 0.7454 | 0.7489 |
| CCCCCCCCCCCCCC | CCCCCCCCCCCC | 0.5629 | 0.4371 | 313 | 0.7431 | 0.7449 |
| CCCCCCCCCCCCCC | CCCCCCCCCCCC | 0.3640 | 0.6360 | 313 | 0.7397 | 0.7420 |
| CCCCCCCCCCCCCC | CCCCCCCCCCCC | 0.1767 | 0.8233 | 313 | 0.7370 | 0.7406 |
| CCCCCCCCCCCCCC | CCCCCCCCCCCC | 0 | 1 | 313 | 0.7347 | 0.7343 |

**Table S11 Comparison of experimental and predicted density values for the hydrocarbon binary mixture dataset using the general model.**

| SMILES_1 | SMILES_2 | x1 | x2 | $T$(K) | $\rho_{exp}$ (g/cm$^3$) | $\rho_{pred}$ (g/cm$^3$) |
|---|---|---|---|---|---|---|
| CCCCCCCCCCCCCC | CCCCCCCCCCCCCCC=C | 1 | 0 | 298 | 0.7603 | 0.7614 |
| CCCCCCCCCCCCCC | CCCCCCCCCCCCCCC=C | 0.8190 | 0.1810 | 298 | 0.7625 | 0.7644 |
| CCCCCCCCCCCCCC | CCCCCCCCCCCCCCC=C | 0.6292 | 0.3708 | 298 | 0.7665 | 0.7667 |
| CCCCCCCCCCCCCC | CCCCCCCCCCCCCCC=C | 0.4299 | 0.5701 | 298 | 0.7702 | 0.7694 |
| CCCCCCCCCCCCCC | CCCCCCCCCCCCCCC=C | 0.2205 | 0.7795 | 298 | 0.7744 | 0.7741 |
| CCCCCCCCCCCCCC | CCCCCCCCCCCCCC=C | 0 | 1 | 298 | 0.7775 | 0.7769 |
| CCCCCCCCCCCCCC | CCCCCCCCCCCC | 1 | 0 | 298 | 0.7603 | 0.7614 |
| CCCCCCCCCCCCCC | CCCCCCCCCCCC | 0.7745 | 0.2255 | 298 | 0.7571 | 0.7597 |
| CCCCCCCCCCCCCC | CCCCCCCCCCCC | 0.5629 | 0.4371 | 298 | 0.7534 | 0.7564 |
| CCCCCCCCCCCCCC | CCCCCCCCCCCC | 0.3640 | 0.6360 | 298 | 0.7506 | 0.7534 |
| CCCCCCCCCCCCCC | CCCCCCCCCCCC | 0.1767 | 0.8233 | 298 | 0.7479 | 0.7500 |
| CCCCCCCCCCCCCC | CCCCCCCCCCCC | 0 | 1 | 298 | 0.7455 | 0.7470 |
| CCCCCCCCCCCCCC | CCCCCCCCCCCCCC=C | 1 | 0 | 313 | 0.7485 | 0.7463 |
| CCCCCCCCCCCCCC | CCCCCCCCCCCCCC=C | 0.8190 | 0.1810 | 313 | 0.7520 | 0.7501 |
| CCCCCCCCCCCCCC | CCCCCCCCCCCCCC=C | 0.6292 | 0.3708 | 313 | 0.7556 | 0.7531 |
| CCCCCCCCCCCCCC | CCCCCCCCCCCCCC=C | 0.4299 | 0.5701 | 313 | 0.7593 | 0.7562 |
| CCCCCCCCCCCCCC | CCCCCCCCCCCCCC=C | 0.2205 | 0.7795 | 313 | 0.7627 | 0.7606 |
| CCCCCCCCCCCCCC | CCCCCCCCCCCCCC=C | 0 | 1 | 313 | 0.7667 | 0.7629 |
| CCCCCCCCCCCCCC | CCCCCCCCCCCC | 1 | 0 | 313 | 0.7485 | 0.7463 |
| CCCCCCCCCCCCCC | CCCCCCCCCCCC | 0.7745 | 0.2255 | 313 | 0.7454 | 0.7450 |
| CCCCCCCCCCCCCC | CCCCCCCCCCCC | 0.5629 | 0.4371 | 313 | 0.7431 | 0.7421 |
| CCCCCCCCCCCCCC | CCCCCCCCCCCC | 0.3640 | 0.6360 | 313 | 0.7397 | 0.7391 |
| CCCCCCCCCCCCCC | CCCCCCCCCCCC | 0.1767 | 0.8233 | 313 | 0.7370 | 0.7355 |
| CCCCCCCCCCCCCC | CCCCCCCCCCCC | 0 | 1 | 313 | 0.7347 | 0.7322 |

**Table S12 Comparison of experimental and predicted density values for the ester and anhydride binary mixture dataset using the ester & anhydride-specific model.**

| SMILES_1 | SMILES_2 | x1 | x2 | $T$(K) | $\rho_{exp}$ (g/cm$^3$) | $\rho_{pred}$ (g/cm$^3$) |
|---|---|---|---|---|---|---|
| CCCCCCCCCCCCOC(=O)C(=C)C | CC(C)CCCCC(=O)OCCCCCCCCCC | 1 | 0 | 298 | 0.8672 | 0.8627 |
| CCCCCCCCCCCCOC(=O)C(=C)C | CC(C)CCCCC(=O)OCCCCCCCCCC | 0.8173 | 0.1827 | 298 | 0.8857 | 0.8827 |
| CCCCCCCCCCCCOC(=O)C(=C)C | CC(C)CCCCC(=O)OCCCCCCCCCC | 0.6265 | 0.3735 | 298 | 0.8768 | 0.8678 |
| CCCCCCCCCCCCOC(=O)C(=C)C | CC(C)CCCCC(=O)OCCCCCCCCCC | 0.4271 | 0.5729 | 298 | 0.8779 | 0.8686 |
| CCCCCCCCCCCCOC(=O)C(=C)C | CC(C)CCCCC(=O)OCCCCCCCCCC | 0.2185 | 0.7815 | 298 | 0.8846 | 0.8826 |
| CCCCCCCCCCCCOC(=O)C(=C)C | CC(C)CCCCC(=O)OCCCCCCCCCC | 0 | 1 | 298 | 0.8698 | 0.8613 |
| CCCCCCCCCCCCOC(=O)C(=C)C | CC(C)CCCCC(=O)OCCCCCCCCCC | 1 | 0 | 313 | 0.8763 | 0.8645 |
| CCCCCCCCCCCCOC(=O)C(=C)C | CC(C)CCCCC(=O)OCCCCCCCCCC | 0.8173 | 0.1827 | 313 | 0.8537 | 0.8525 |
| CCCCCCCCCCCCOC(=O)C(=C)C | CC(C)CCCCC(=O)OCCCCCCCCCC | 0.6265 | 0.3735 | 313 | 0.8067 | 0.8090 |
| CCCCCCCCCCCCOC(=O)C(=C)C | CC(C)CCCCC(=O)OCCCCCCCCCC | 0.4271 | 0.5729 | 313 | 0.8065 | 0.8100 |
| CCCCCCCCCCCCOC(=O)C(=C)C | CC(C)CCCCC(=O)OCCCCCCCCCC | 0.2185 | 0.7815 | 313 | 0.8430 | 0.8530 |
| CCCCCCCCCCCCOC(=O)C(=C)C | CC(C)CCCCC(=O)OCCCCCCCCCC | 0 | 1 | 313 | 0.8695 | 0.8539 |
| CCCCC(CC)C(=O)OCC(COC(=O)C(CC)CCCC)(COC(=O)C(CC)CCCC)COC(=O)C(CC)CCCC | CCCCCCOC(=O)C(=C)C | 0.0623 | 0.9377 | 298 | 0.8953 | 0.8852 |
| CCCCC(CC)C(=O)OCC(COC(=O)C(CC)CCCC)(COC(=O)C(CC)CCCC)COC(=O)C(CC)CCCC | CCCCCCOC(=O)C(=C)C | 0 | 1 | 298 | 0.8815 | 0.8792 |

**Table S13 Comparison of experimental and predicted density values for the ester and anhydride binary mixture dataset using the general model.**

| SMILES_1 | SMILES_2 | x1 | x2 | *T*(K) | $\rho_{exp}$ (g/cm$^3$) | $\rho_{pred}$ (g/cm$^3$) |
|---|---|---|---|---|---|---|
| CCCCCCCCCCCCCOC(=O)C(=C)C | CC(C)CCCCC(=O)OCCCCCCCCCC | 1 | 0 | 298 | 0.8672 | 0.8677 |
| CCCCCCCCCCCCCOC(=O)C(=C)C | CC(C)CCCCC(=O)OCCCCCCCCCC | 0.8173 | 0.1827 | 298 | 0.8857 | 0.8822 |
| CCCCCCCCCCCCCOC(=O)C(=C)C | CC(C)CCCCC(=O)OCCCCCCCCCC | 0.6265 | 0.3735 | 298 | 0.8768 | 0.8716 |
| CCCCCCCCCCCCCOC(=O)C(=C)C | CC(C)CCCCC(=O)OCCCCCCCCCC | 0.4271 | 0.5729 | 298 | 0.8779 | 0.8706 |
| CCCCCCCCCCCCCOC(=O)C(=C)C | CC(C)CCCCC(=O)OCCCCCCCCCC | 0.2185 | 0.7815 | 298 | 0.8846 | 0.8808 |
| CCCCCCCCCCCCCOC(=O)C(=C)C | CC(C)CCCCC(=O)OCCCCCCCCCC | 0 | 1 | 298 | 0.8698 | 0.8693 |
| CCCCCCCCCCCCCOC(=O)C(=C)C | CC(C)CCCCC(=O)OCCCCCCCCCC | 1 | 0 | 313 | 0.8763 | 0.8706 |
| CCCCCCCCCCCCCOC(=O)C(=C)C | CC(C)CCCCC(=O)OCCCCCCCCCC | 0.8173 | 0.1827 | 313 | 0.8537 | 0.8575 |
| CCCCCCCCCCCCCOC(=O)C(=C)C | CC(C)CCCCC(=O)OCCCCCCCCCC | 0.6265 | 0.3735 | 313 | 0.8067 | 0.8070 |
| CCCCCCCCCCCCCOC(=O)C(=C)C | CC(C)CCCCC(=O)OCCCCCCCCCC | 0.4271 | 0.5729 | 313 | 0.8065 | 0.8069 |
| CCCCCCCCCCCCCOC(=O)C(=C)C | CC(C)CCCCC(=O)OCCCCCCCCCC | 0.2185 | 0.7815 | 313 | 0.8430 | 0.8485 |
| CCCCCCCCCCCCCOC(=O)C(=C)C | CC(C)CCCCC(=O)OCCCCCCCCCC | 0 | 1 | 313 | 0.8695 | 0.8625 |
| CCCCC(CC)C(=O)OCC(COC(=O)C(CC)CCCC)(COC(=O)C(CC)CCCC)COC(=O)C(CC)CCCC | CCCCCCOC(=O)C(=C)C | 0.0623 | 0.9377 | 298 | 0.8953 | 0.9073 |
| CCCCC(CC)C(=O)OCC(COC(=O)C(CC)CCCC)(COC(=O)C(CC)CCCC)COC(=O)C(CC)CCCC | CCCCCCOC(=O)C(=C)C | 0 | 1 | 298 | 0.8815 | 0.8847 |

# Prediction performance of viscosity and density for ternary hydrocarbon mixtures.

**Table S14 Prediction performance of dynamic viscosity for ternary hydrocarbon mixtures.**

| SMILES_1 | SMILES_2 | SMILES_3 | x1 | x2 | x3 | *T*(K) | $\eta_{exp}$(cP) | $\eta_{pred}$ (cP) |
|---|---|---|---|---|---|---|---|---|
| CCCCCCCCCCCC | CCCCCCCCCCCCCC | CCCCCCCCCCCCCCC=C | 0.2361 | 0.4055 | 0.3584 | 298 | 1.7811 | 1.6756 |
| CCCCCCCCCCCC | CCCCCCCCCCCCCC | CCCCCCCCCCCCCCC=C | 0.4468 | 0.3836 | 0.1696 | 298 | 1.6071 | 1.4957 |
| CCCCCCCCCCCC | CCCCCCCCCCCCCC | CCCCCCCCCCCCCCC=C | 0.4570 | 0.1962 | 0.3468 | 298 | 1.7835 | 1.6031 |
| CCCCCCCCCCCC | CCCCCCCCCCCCCC | CCCCCCCCCCCCCCC=C | 0.2361 | 0.4055 | 0.3584 | 313 | 1.3855 | 1.2738 |
| CCCCCCCCCCCC | CCCCCCCCCCCCCC | CCCCCCCCCCCCCCC=C | 0.4468 | 0.3836 | 0.1696 | 313 | 1.2804 | 1.1590 |
| CCCCCCCCCCCC | CCCCCCCCCCCCCC | CCCCCCCCCCCCCCC=C | 0.4570 | 0.1962 | 0.3468 | 313 | 1.3698 | 1.2241 |
| CCCCCCCCCCCCCC | CCCCCCCCCCCCCCC=C | CCCCCCCCCCCCCCCCC=C | 0.2304 | 0.4074 | 0.3621 | 298 | 2.7800 | 2.1429 |
| CCCCCCCCCCCCCC | CCCCCCCCCCCCCCC=C | CCCCCCCCCCCCCCCCC=C | 0.4392 | 0.3882 | 0.1726 | 298 | 2.5111 | 2.0043 |
| CCCCCCCCCCCCCC | CCCCCCCCCCCCCCC=C | CCCCCCCCCCCCCCCCC=C | 0.4489 | 0.1984 | 0.3527 | 298 | 2.6322 | 2.0462 |
| CCCCCCCCCCCCCC | CCCCCCCCCCCCCCC=C | CCCCCCCCCCCCCCCCC=C | 0.2304 | 0.4074 | 0.3621 | 313 | 2.0726 | 1.5762 |
| CCCCCCCCCCCCCC | CCCCCCCCCCCCCCC=C | CCCCCCCCCCCCCCCCC=C | 0.4392 | 0.3882 | 0.1726 | 313 | 1.8117 | 1.4738 |
| CCCCCCCCCCCCCC | CCCCCCCCCCCCCCC=C | CCCCCCCCCCCCCCCCC=C | 0.4489 | 0.1984 | 0.3527 | 313 | 1.9461 | 1.5008 |

**Table S15 Prediction performance of density for ternary hydrocarbon mixtures.**

| SMILES_1 | SMILES_2 | SMILES_3 | x1 | x2 | x3 | *T*(K) | $\rho_{exp}$ (g/cm$^3$) | $\rho_{pred}$ (g/cm$^3$) |
|---|---|---|---|---|---|---|---|---|
| CCCCCCCCCCCC | CCCCCCCCCCCCCC | CCCCCCCCCCCCCCC=C | 0.2361 | 0.4055 | 0.3584 | 298 | 0.7618 | 0.7414 |
| CCCCCCCCCCCC | CCCCCCCCCCCCCC | CCCCCCCCCCCCCCC=C | 0.4468 | 0.3836 | 0.1696 | 298 | 0.7552 | 0.7385 |
| CCCCCCCCCCCC | CCCCCCCCCCCCCC | CCCCCCCCCCCCCCC=C | 0.4570 | 0.1962 | 0.3468 | 298 | 0.7583 | 0.7405 |
| CCCCCCCCCCCC | CCCCCCCCCCCCCC | CCCCCCCCCCCCCCC=C | 0.2361 | 0.4055 | 0.3584 | 313 | 0.7512 | 0.7356 |
| CCCCCCCCCCCC | CCCCCCCCCCCCCC | CCCCCCCCCCCCCCC=C | 0.4468 | 0.3836 | 0.1696 | 313 | 0.7443 | 0.7326 |
| CCCCCCCCCCCC | CCCCCCCCCCCCCC | CCCCCCCCCCCCCCC=C | 0.4570 | 0.1962 | 0.3468 | 313 | 0.7486 | 0.7350 |
| CCCCCCCCCCCCCC | CCCCCCCCCCCCCCC=C | CCCCCCCCCCCCCCCCC=C | 0.2304 | 0.4074 | 0.3621 | 298 | 0.7768 | 0.7468 |
| CCCCCCCCCCCCCC | CCCCCCCCCCCCCCC=C | CCCCCCCCCCCCCCCCC=C | 0.4392 | 0.3882 | 0.1726 | 298 | 0.7726 | 0.7441 |
| CCCCCCCCCCCCCC | CCCCCCCCCCCCCCC=C | CCCCCCCCCCCCCCCCC=C | 0.4489 | 0.1984 | 0.3527 | 298 | 0.7733 | 0.7459 |
| CCCCCCCCCCCCCC | CCCCCCCCCCCCCCC=C | CCCCCCCCCCCCCCCCC=C | 0.2304 | 0.4074 | 0.3621 | 313 | 0.7673 | 0.7411 |
| CCCCCCCCCCCCCC | CCCCCCCCCCCCCCC=C | CCCCCCCCCCCCCCCCC=C | 0.4392 | 0.3882 | 0.1726 | 313 | 0.7615 | 0.7384 |
| CCCCCCCCCCCCCC | CCCCCCCCCCCCCCC=C | CCCCCCCCCCCCCCCCC=C | 0.4489 | 0.1984 | 0.3527 | 313 | 0.7631 | 0.7407 |

**Figure S1. Molecular structures of the five representative hydrocarbon components (H1–H5) and the C30 PAO model molecule used in Table S16 below.** H1 is an isoparaffin, H2 is an alkylcyclohexane, H3 is an alkylbenzene, H4 is an alkyldecalin, and H5 is an alkylnaphthalene. All five components are paired with the same PAO molecule to form binary hydrocarbon + PAO mixtures.

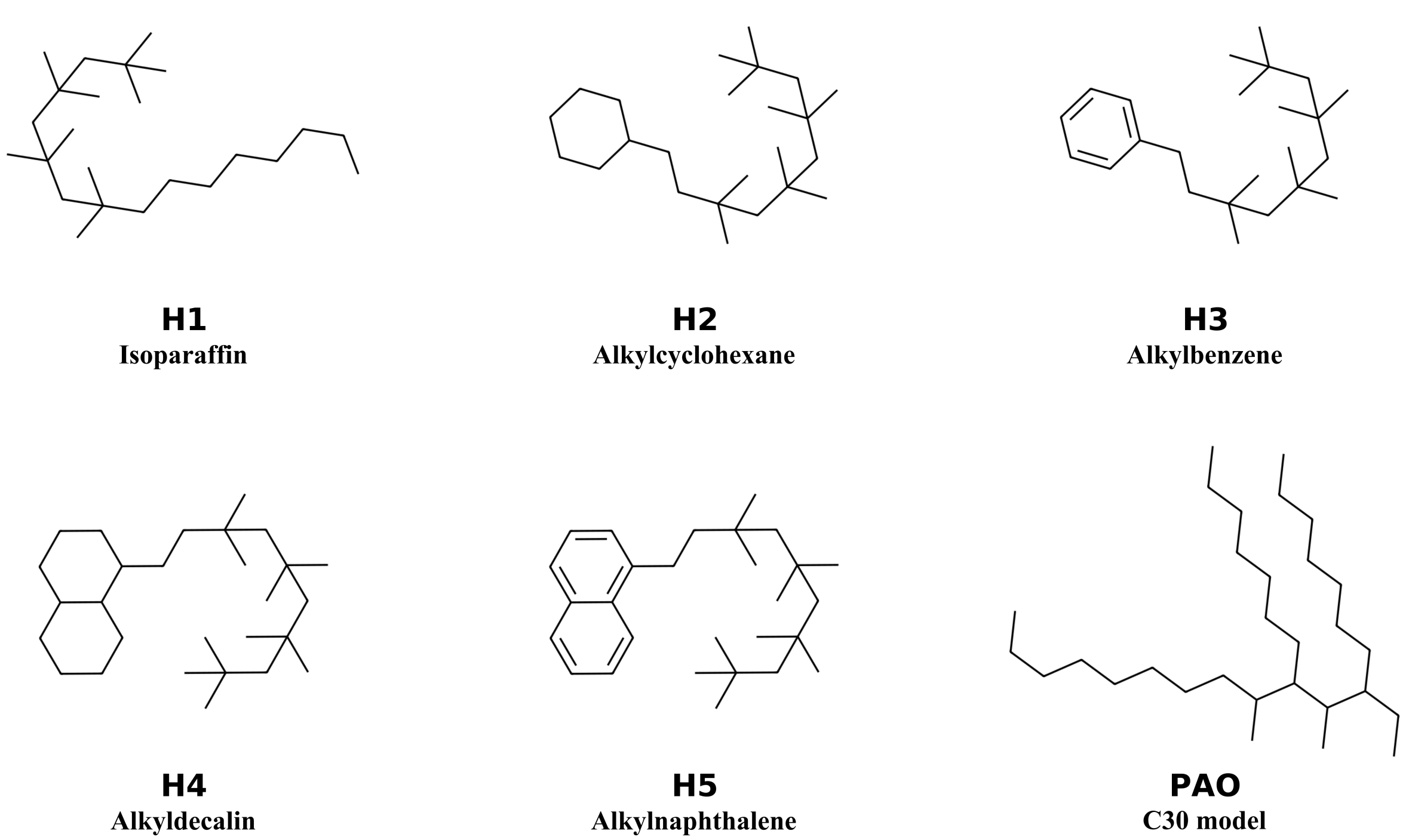

**Table S16. Predicted dynamic viscosity values for the pure component and four binary mixtures at various temperatures.**

**(a) Pure PAO reference (group 0)**

| SMILES | X | T (K) | ν(cSt) |
|---|---|---|---|
| CCCCCCCC(CC)C(C)C(CCCCCCC)C(C)CCCCCCCC | 1.0 | 313.15 | 13.485 |
| CCCCCCCC(CC)C(C)C(CCCCCCC)C(C)CCCCCCCC | 1.0 | 373.15 | 4.960 |

**(b) PAO + H1 (Isoparaffin)**

**Component 1: CCCCCCCC(CC)C(C)C(CCCCCCC)C(C)CCCCCCCC**

**Component 2: CCCCCCCCC(C)(C)CC(C)(C)CC(C)(C)CC(C)(C)C**

| $x_1$ | $x_2$ | $\nu_{40}$ (cSt) | $\nu_{100}$ (cSt) | $\nu_{40}/\nu_{100}$ |
|---|---|---|---|---|
| 0.98 | 0.02 | 13.109 | 4.819 | 2.720 |
| 0.96 | 0.04 | 12.733 | 4.676 | 2.723 |
| 0.94 | 0.06 | 12.327 | 4.522 | 2.726 |
| 0.92 | 0.08 | 11.912 | 4.367 | 2.727 |
| 0.9 | 0.1 | 11.548 | 4.233 | 2.728 |
| 0.88 | 0.12 | 11.220 | 4.107 | 2.732 |
| 0.86 | 0.14 | 10.935 | 3.995 | 2.737 |
| 0.84 | 0.16 | 10.674 | 3.892 | 2.743 |
| 0.82 | 0.18 | 10.435 | 3.796 | 2.749 |
| 0.8 | 0.2 | 10.205 | 3.701 | 2.757 |

**(c) PAO + H2 (Alkylcyclohexane)**

**Component 1: CCCCCCCC(CC)C(C)C(CCCCCCC)C(C)CCCCCCCC**

**Component 2: C1CCCCC1CCC(C)(C)CC(C)(C)CC(C)(C)CC(C)(C)C**

| $x_1$ | $x_2$ | $\nu_{40}$ (cSt) | $\nu_{100}$ (cSt) | $\nu_{40}/\nu_{100}$ |
|---|---|---|---|---|
| 0.98 | 0.02 | 13.145 | 4.815 | 2.730 |
| 0.96 | 0.04 | 12.817 | 4.677 | 2.740 |
| 0.94 | 0.06 | 12.468 | 4.549 | 2.741 |
| 0.92 | 0.08 | 12.108 | 4.416 | 2.742 |
| 0.9 | 0.1 | 11.786 | 4.291 | 2.747 |
| 0.88 | 0.12 | 11.520 | 4.179 | 2.756 |
| 0.86 | 0.14 | 11.289 | 4.078 | 2.768 |
| 0.84 | 0.16 | 11.090 | 3.985 | 2.783 |
| 0.82 | 0.18 | 10.916 | 3.895 | 2.802 |
| 0.8 | 0.2 | 10.762 | 3.807 | 2.827 |

**(d) PAO + H3 (Alkylbenzene)**

**Component 1: CCCCCCCC(CC)C(C)C(CCCCCCC)C(C)CCCCCCCC**

**Component 2: c1ccccc1CCC(C)(C)CC(C)(C)CC(C)(C)CC(C)(C)C**

| $x_1$ | $x_2$ | $\nu_{40}$ (cSt) | $\nu_{100}$ (cSt) | $\nu_{40}/\nu_{100}$ |
|---|---|---|---|---|
| 0.98 | 0.02 | 13.107 | 4.807 | 2.727 |
| 0.96 | 0.04 | 12.740 | 4.661 | 2.733 |
| 0.94 | 0.06 | 12.331 | 4.509 | 2.734 |
| 0.92 | 0.08 | 11.914 | 4.356 | 2.735 |
| 0.9 | 0.1 | 11.557 | 4.224 | 2.736 |
| 0.88 | 0.12 | 11.265 | 4.110 | 2.741 |
| 0.86 | 0.14 | 10.974 | 3.996 | 2.746 |
| 0.84 | 0.16 | 10.713 | 3.892 | 2.752 |
| 0.82 | 0.18 | 10.492 | 3.802 | 2.759 |
| 0.8 | 0.2 | 10.286 | 3.717 | 2.767 |

**(e) PAO + H4 (Alkyldecalin)**

**Component 1: CCCCCCCC(CC)C(C)C(CCCCCCC)C(C)CCCCCCCC**

**Component 2: C1CCC2CCCCC2C1CCC(C)(C)CC(C)(C)CC(C)(C)CC(C)(C)C**

| $x_1$ | $x_2$ | $\nu_{40}$ (cSt) | $\nu_{100}$ (cSt) | $\nu_{40}/\nu_{100}$ |
|---|---|---|---|---|
| 0.98 | 0.02 | 13.198 | 4.825 | 2.735 |
| 0.96 | 0.04 | 12.934 | 4.698 | 2.753 |
| 0.94 | 0.06 | 12.640 | 4.581 | 2.759 |
| 0.92 | 0.08 | 12.398 | 4.463 | 2.778 |
| 0.9 | 0.1 | 12.197 | 4.355 | 2.801 |
| 0.88 | 0.12 | 12.022 | 4.255 | 2.826 |
| 0.86 | 0.14 | 11.876 | 4.157 | 2.857 |
| 0.84 | 0.16 | 11.752 | 4.067 | 2.890 |
| 0.82 | 0.18 | 11.642 | 3.981 | 2.925 |
| 0.8 | 0.2 | 11.552 | 3.899 | 2.963 |

**(f) PAO + H5 (Alkylnaphthalene)**

**Component 1: CCCCCCCC(CC)C(C)C(CCCCCCC)C(C)CCCCCCCC**

**Component 2: c1ccc2ccccc2c1CCC(C)(C)CC(C)(C)CC(C)(C)CC(C)(C)C**

| $x_1$ | $x_2$ | $\nu_{40}$ (cSt) | $\nu_{100}$ (cSt) | $\nu_{40}/\nu_{100}$ |
|---|---|---|---|---|
| 0.98 | 0.02 | 13.179 | 4.822 | 2.733 |
| 0.96 | 0.04 | 12.890 | 4.692 | 2.747 |
| 0.94 | 0.06 | 12.598 | 4.571 | 2.756 |
| 0.92 | 0.08 | 12.312 | 4.452 | 2.765 |
| 0.9 | 0.1 | 12.067 | 4.340 | 2.780 |
| 0.88 | 0.12 | 11.847 | 4.236 | 2.797 |
| 0.86 | 0.14 | 11.656 | 4.133 | 2.820 |
| 0.84 | 0.16 | 11.490 | 4.037 | 2.846 |
| 0.82 | 0.18 | 11.342 | 3.945 | 2.875 |
| 0.8 | 0.2 | 11.217 | 3.860 | 2.906 |